\documentclass[twocolumn,tighten]{aastex61}

\usepackage{amsmath,amstext}
\usepackage{graphicx}
\usepackage{color}

\received{MMMMM DD, YYYY}
\revised{MMMMM DD, YYYY}
\accepted{MMMMM DD, YYYY}

\submitjournal{ApJ}

\shorttitle{Testing Isotropic Universe}
\shortauthors{\v{R}\'{\i}pa et al.}

\begin{document}

\title{Testing Isotropic Universe Using the Gamma-Ray Burst \\
Data of {\em Fermi} / GBM}

\author{Jakub \v{R}\'{\i}pa}
\email{jripa@ntu.edu.tw}
\affiliation{National Taiwan University,\\
Leung Center for Cosmology and Particle Astrophysics,\\
No.1, Sec.4, Roosevelt Road, Taipei 10617, Taiwan (R.O.C)}

\author{Arman Shafieloo}
\email{shafieloo@kasi.re.kr}
\affiliation{Korea Astronomy and Space Science Institute,\\
Daejeon 305-348, Korea}
\affiliation{University of Science and Technology,\\
Daejeon 34113, Korea}

\begin{abstract}
The sky distribution of Gamma-Ray Bursts (GRBs) has been intensively studied by various groups
for more than two decades. Most of these studies test the isotropy of GRBs based on their sky
number density distribution. In this work we propose an approach to test the isotropy of the
Universe through inspecting the isotropy of the properties of GRBs such as their duration,
fluences and peak fluxes at various energy bands and different time scales. We apply this
method on the {\em Fermi} / Gamma-ray Burst Monitor (GBM) data sample containing 1591 GRBs.
The most noticeable feature we found is near the Galactic coordinates
$l\approx 30^\circ$, $b\approx 15^\circ$ and radius $r\approx 20^\circ-40^\circ$.
The inferred probability for the occurrence of such an anisotropic signal (in a random isotropic sample) is derived to be less than a  percent in some of the tests while the other tests give results consistent with isotropy.
These are based on the comparison of the results from the real data with the randomly shuffled
data samples.  Considering large number of statistics we used in this work (which some of them are correlated to each other) we can anticipate that the detected feature could be result of statistical fluctuations. Moreover, we noticed a considerably low number of GRBs in this particular
patch which might be due to some instrumentation or observational effects that can consequently
affect our statistics through some systematics. Further investigation is highly desirable in order clarify about this result, e.g. utilizing a larger future {\em Fermi} / GBM data sample as well as data samples of other GRB missions and also looking for possible systematics.

\end{abstract}


\section{Introduction}
\label{sec:intro}

Recent observations claimed the existence of large-scale structures in the Universe of sizes of several
hundreds of megaparsecs or even beyond one gigaparsec. For example the Sloan Great Wall of galaxies
\citep{got05} 419\,Mpc long at redshift $z=0.073$ has been reported. The radio observations of the
National Radio Astronomy Observatory's Very Large Array Sky Survey suggested a 140\,Mpc completely
empty void at $z \leq 1$ \citep{rud07}. Or a Huge Large Quasar Group (Huge-LQG) of longest
dimension $\sim 1240$\,Mpc at mean redshift $\bar z=1.27$ has been claimed \citep{clo13}.
However, \citet{nad13} argued that interpreting the Huge-LQG as a structure is questionable.

Concerning the Gamma-Ray Bursts (GRBs) \citep{ved09,kou12,kum15}, initially they had been claimed to be
distributed isotropically on the sky \citep{mee92,bri96,teg96}. Later several works claimed that short GRBs
($T_{90}<2$\,s) were distributed anisotropically
\citep{bal98,bal99,mag03,vav08,tar17},
where $T_{90}$ is the duration during which 90\,\% of the detected counts
from a GRB is accumulated \citep{kou93}.
However, \citet{mes00a,mes00b,lit01,ber08,ver10a,ukw16} claimed different results.
The pertinent intermediate-duration GRBs (2\,s $\lesssim T_{90} \lesssim$ 10\,s)
\citep{muk98,hor98,blg01,hor02,ho08,hor09,huj09,rip09,ho10,ver10b,uga11,rip12,zit15,hor16,
rip16,yan16,kul17} were found to be distributed anisotropically on the sky
\citep{mes00a,mes00b,lit01,vav08,ver10a}.
However, \citet{mes03,ukw16} claimed different results.
It was also found that very short GRBs ($T_{90} \leq 100$\,ms) were distributed anisotropically
\citep{cli05,ukw16}.
Long GRBs ($T_{90}>2$\,s) were proclaimed to be distributed isotropically
\citep{bal98,bal99,mes00a,mes00b,mag03,vav08,tar17,ukw16}, however
\citet{mes03} claimed different result.
For review of these works see \citet{mes09a,mes09b,mes17}.

Recently, \citet{ho14,ho15} studied the spatial distribution of GRBs taking into account their
redshift. They concluded that there was a statistically significant clustering of GRBs at
redshift $1.6 < z \leq 2.1$ and that the size of the structure defined by these GRBs was about
2\,000--3\,000\,Mpc, so called Hercules--Corona Borealis Great Wall.
Contrary to this, \citet{ukw16} claimed that their redshift-dependent analysis did
not provide evidence of significant clustering. A discussion of the implications of the results
obtained by \citet{ho14} on the Cosmological Principle can be found in \citet{li15}.

\citet{bal15} reported the discovery of a giant ring-like structure with a diameter of
1\,720 Mpc, displayed by 9 GRBs. The ring has a major diameter of $43^\circ$ and a minor
diameter of $30^\circ$ at a distance of 2\,770 Mpc in the redshift range of $0.78 < z <0.86$.

Another work testing the isotropic distribution of GRBs and making a comparison with
Cosmic Microwave Background (CMB) can be found in \citet{gan16}.

Further information about the tests of the isotropy or homogeneity of the Universe
using SNe Ia, CMB, dark energy, galaxies and large-scale structures see, e.g.
\citet{hin96,fri05a,fri05b,hog05,cam11,col11,scr12,Feindt:2013pma,app14,fer14,pla14,app15,jav15,pla16,che17,jav17}
and references therein.

All these studies, which use GRB data, test the isotropy based on the distribution of
the number densities of GRBs. On the other hand, in this paper we test the isotropy of the
Universe through inspecting the isotropy of the properties of GRBs such as their duration,
fluences and peak fluxes at various energy bands and different time scales.

The paper is organized as follows. Section~\ref{sec:sample} describes the used data sample.
Section~\ref{sec:method} details the proposed methodology. The results are presented in
Section~\ref{sec:results} and in Appendix~\ref{sec:append_tables}. Our findings are discussed
in Section~\ref{sec:discuss} and Section~\ref{sec:conclude} summarizes the conclusions.

\section{Data Sample}
\label{sec:sample}

The database of the {\em Fermi} satellite\footnote{http://fermi.gsfc.nasa.gov/} \citep{atw94} and
particularly of the Gamma-ray Burst Monitor (GBM) instrument \citep{mee09} has
been utilized in this work. Specifically we employed the FERMIGBRST - Fermi GBM Burst
Catalog\footnote{https://heasarc.gsfc.nasa.gov/W3Browse/fermi/fermigbrst.html}
\citep{gru14,vki14,bha16}, which is being constantly updated. While methods for measuring GRB properties more precisely has been suggested \citep{sze13}, and
alternative data mining strategies has been developed to identify non-triggered events \citep{bag16}, this is still one of the most complete burst catalog to date.
A sample containing 1594 GRBs with the first and the last event detected on
14 Jul 2008 and on 15 Apr 2015, respectively, is used. Only the following
quantities contained in the catalog are applied in our analysis:

\begin{itemize}
\item ``LII'' is the galactic longitude -- denoted as $l$ ($^{\circ}$).
\item ``BII'' is the galactic latitude -- denoted as $b$ ($^{\circ}$).
\item ``Error\_Radius'' is the uncertainty in the position ($^{\circ}$).
\item ``T90'' is the duration during which 90\,\% of the burst's fluence was accumulated nominally
in the $(50-300)$\,keV energy range -- in what follows denoted as $T_{90}$ (s).
The start and the end of the $T_{90}$ interval is defined by the time at which 5\,\% and 95\,\%
of the total fluence have been detected, respectively.
\item ``Flux\_64'', ``Flux\_256'', and ``Flux\_1024'' are the peak fluxes on the 64-ms, 256-ms, and
1024-ms timescales, nominally in the energy range of $(10-1000)$\,keV - here denoted as $F_{64}$,
$F_{256}$, and $F_{1024}$, respectively.
\item ``Flux\_BATSE\_64'', ``Flux\_BATSE\_256'', and \\
``Flux\_BATSE\_1024'' are the peak fluxes on
the 64-ms, 256-ms, and 1024-ms timescales, in the BATSE standard $(50-300)$\,keV energy band -- here denoted
as $F_{64\mathrm{,B}}$, $F_{256\mathrm{,B}}$, and $F_{1024\mathrm{,B}}$, respectively.
\item ``Fluence'' is the time integrated flux over the whole burst's duration nominally in the
$(10-1000)$\,keV energy range -- here denoted as $S$.
\item ``Fluence\_BATSE'' is the time integrated flux over the whole burst's duration in the BATSE
standard $(50-300)$\,keV energy band -- here denoted as $S_{\mathrm{B}}$.
\end{itemize}

All peak fluxes have units of ph\,cm$^{-2}$\,s$^{-1}$ and both fluences have units of erg\,cm$^{-2}$.
In one case, GRB150120123, the energy range used for measuring $T_{90}$ was $(300-500)$\,keV.
In two cases the energy ranges used for deriving peak fluxes $F_{64}$, $F_{256}$, $F_{1024}$ and
fluence $S$ were $(30-500)$\,keV for GRB100918863 and $(0.01-100)$\,keV for GRB110213876.

In this database all 1594 GRBs have measured galactic coordinates.
The number of GRBs in this database with measured galactic coordinates and $T_{90}$, 
$F_{64}$, $F_{256}$, $F_{1024}$, $F_{64\mathrm{,B}}$, $F_{256\mathrm{,B}}$,
$F_{1024\mathrm{,B}}$, $S$, and $S_{\mathrm{B}}$ is 1591.
These are the tested quantities in our analysis. Our only selection criterion is that the galactic
coordinates and the tested quantities must be measured. Thus the data samples used in our analysis
have sizes of 1591 events.

\section{Method}
\label{sec:method}

In order to test the isotropy of the universe we propose following approach.
We do not study the number density of GRBs on the sky, but we test the isotropy by
analyzing the properties of GRBs as they are distributed on the sky.
We compare distributions of a given measured GRB property (e.g. duration or flux)
for a large number of randomly spread patches on the sky with a distribution of the
same GRB property for the whole sky. This approach is based on the principles of the Crossing statistic that has been used in 
different contexts in cosmology\citep{Crossing1,col11,Crossing2,Crossing3,Crossing4}. 

We use several test statistics to give us the
measure of the differences between the distribution for a random patch and the whole sky.
Then we compare the obtained distributions of the test statistics derived from the
measured data with the distributions of the test statistics for randomly shuffled
data to infer the significance of potential anisotropies.
The comparison of the measured data with the randomly shuffled once is the
key step in our method because it means in essence comparison of the measured data
to the isotropically distributed hypothetical sample.
The instrumental sky exposure map is not needed in our work.
The reason is that if in any direction on the sky the number of observed GRBs is reduced
due to the lower exposure time, then the distribution of a given GRB property is known
with lower resolution, but the shape of the distribution should still remain similar.

The null hypothesis is that universe is isotropical. We are testing against this null
hypothesis. More detailed steps of our method are following:
\begin{enumerate}

\begin{figure}[t]
\includegraphics[width=0.45\textwidth]{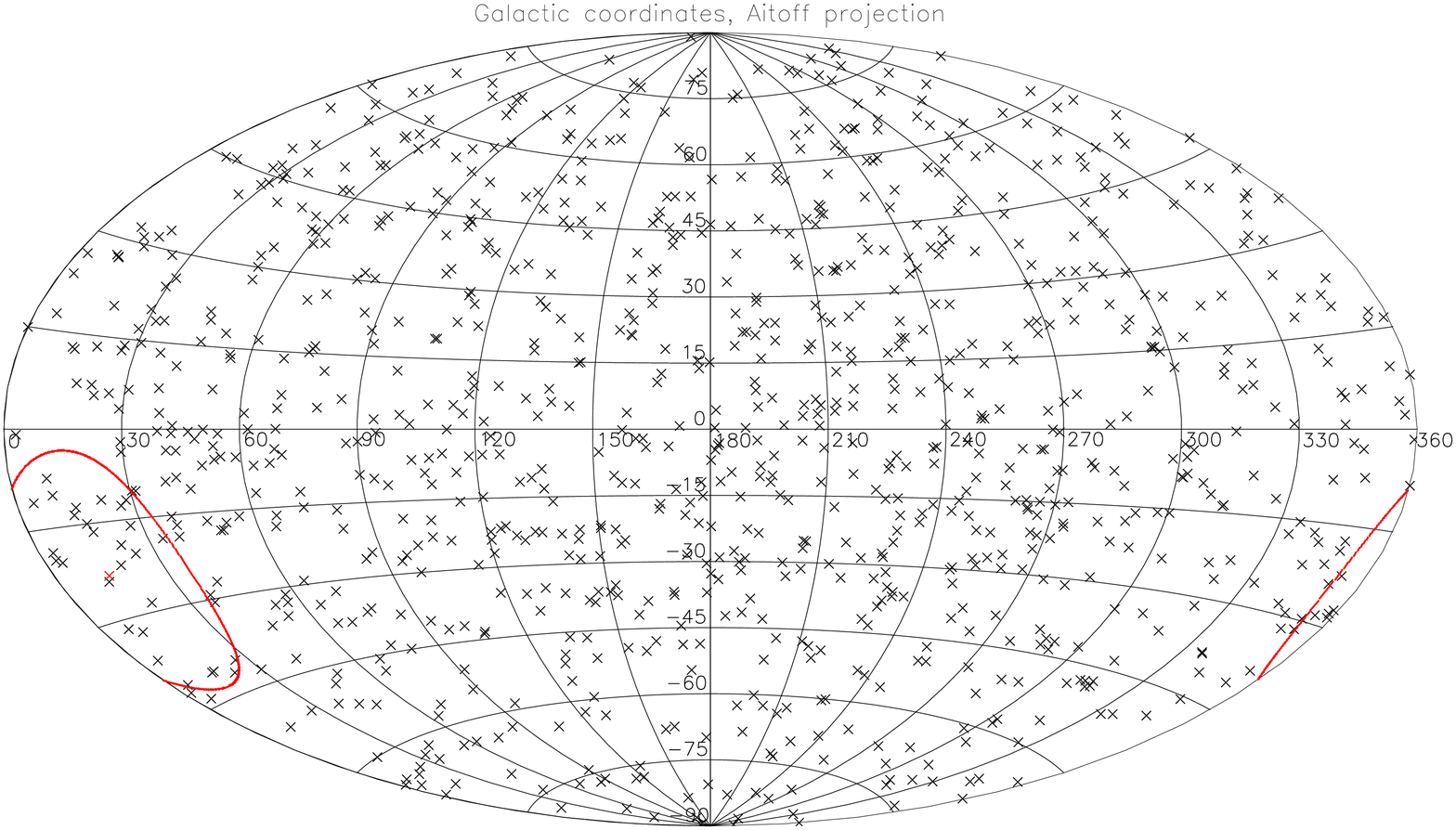}
\caption{The centres of 1000 patches randomly distributed on the sky used in this
work (crosses). The red curve denotes the boundary of a random example patch of radius
$r=20^\circ$ (red cross is its center).}
\label{fig:method_1}
\end{figure}

\item Take the empirical distribution function $F_\mathrm{p}(x)$ of a given tested
quantity $x$ of GRBs, e.g. $x=T_{90}$, $S$, $F$, etc. (see Section~\ref{sec:sample}),
for the whole sky. Number of GRBs in the data sample of the whole sky is $p=1591$.

\item Make 1\,000 patches of a given radius $r$ randomly distributed on the sky.
Afterwards we keep the positions of these patches fixed throughout our analysis
(see Fig.~\ref{fig:method_1}).

\item For the GRBs in these 1\,000 patches take the empirical distribution functions
$G_\mathrm{q}(x)$ of the same property $x$. Number of GRBs in each patch is $q$
and it varies patch to patch.

\item For each patch compare the two empirical distribution functions
$G_\mathrm{q}(x)$ and $F_\mathrm{p}(x)$ by calculating several test statistics
$\xi=D$, $V$, $AD$, or $\chi^2$, where
\begin{enumerate}

\item $D$ is the statistic of the two-sample Kolmogo\-rov--Smirnov (K--S) test
\citep{kol33,smi39,pre07} defined as
\begin{equation}
D=\underset{-\infty < x < \infty}{\mathrm{max}}\left | F_\mathrm{p}(x)-G_\mathrm{q}(x) \right |
\end{equation}
Here $G_\mathrm{q}(x)$ and $F_\mathrm{p}(x)$ are the empirical cumulative
distribution functions (see Fig.~\ref{fig:method_2}).
The test statistic is the maximum value of the
absolute difference between the two empirical cumulative distribution functions.
Since the K--S test tends to be most sensitive around the median value
of the distributions and less sensitive on the tails of the distributions we use
also Kuiper's $V$ and Anderson--Darling's $AD$ statistics which
provide increased sensitivity on the tails (\citet{pre07}).

\item $V$ is the statistic of the two-sample Kuiper test \citep{kui60,pre07} defined as
\begin{equation}
\begin{split}
V=D_{+} + D_{-} = \\
\underset{-\infty < x < \infty}{\mathrm{max}}\left 
[ F_\mathrm{p}(x)-G_\mathrm{q}(x) \right ] + \\
\underset{-\infty < x < \infty}{\mathrm{max}}\left [ G_\mathrm{q}(x)-F_\mathrm{p}(x) \right ]
\end{split}
\end{equation}
Here $G_\mathrm{q}(x)$ and $F_\mathrm{p}(x)$ are the empirical cumulative
distribution functions. The Kuiper's test is related to the K--S test.
The $D_{+}$ and $D_{-}$ represent the absolute values of the most positive and
the most negative differences between the two empirical cumulative distribution
functions.

\item $AD$ is the statistic of the two-sample Anderson--Darling (A--D) test
\citep{and52,dar57,pet76} defined as
\begin{equation}
\begin{split}
AD\equiv A^2_\mathrm{pq}=\\
\frac{pq}{s}\int_{-\infty}^{\infty}\frac{[ F_{p}(x) - G_{q}(x)]^2}
{H_{s}(x)[1-H_{s}(x)]}dH_{s}(x),
\end{split}
\end{equation}
where $G_\mathrm{q}(x)$, $F_\mathrm{p}(x)$ are the two empirical distribution functions
being tested and the $H_{s}(x)=[pF_{p}(x)+qG_{q}(x)]/s$ with $s=p+q$ is the empirical
distribution function of the pooled sample. The above integrand is defined to be zero
whenever $H_{s}(x)=1$. For details see \citet{sch87}.

\item $\chi^2$ is the statistic of the two-sample Chi-square test \citep{pre07}.
In case of $\chi^2$ the frequencies in the binned data of two samples are compared
instead of the empirical distribution functions. The binning for both data sets,
which are being compared, should be the same. We used $k=10$ number of bins and for
this test we binned the logarithmic values: $\log x = \log T_{90}$, $\log S$,
$\log F$, etc.. The $\chi^2$ is defined as
\begin{equation}
\chi^2=\sum_{i=1}^{k}\frac{\left ( K_1R_i-K_2Q_i \right )^2}{R_i+Q_i},
\end{equation}
where $k$ is the number of bins, $R_i$ is the observed frequency for bin $i$ for the first
sample, and $Q_i$ is the observed frequency for bin $i$ for the second sample
(see Fig.~\ref{fig:method_3}). $K_1$ and $K_2$ are the weights used to adjust for unequal
sample sizes
\begin{equation}
K_1=\sqrt{\frac{\sum_{i=1}^{k}Q_i}{\sum_{i=1}^{k}R_i}}
\end{equation}
and
\begin{equation}
K_2=\sqrt{\frac{\sum_{i=1}^{k}R_i}{\sum_{i=1}^{k}Q_i}}.
\end{equation}
\end{enumerate}

\item This gives us a distribution of 1\,000 values of $\xi^\mathrm{m}$ obtained from the
measured data sample. In what follows the superscript $m$ used with any quantity, means that
the quantity is related to the actual measured data sample.

\item In the next step we randomly shuffle our measured data sample. Each measurement is
represented by a triplet $\{l_{\mathrm{i}}$, $b_{\mathrm{i}}$, $x_{\mathrm{i}}\}$. We randomly
shuffle $x_{\mathrm{i}}$ keeping $l_{\mathrm{i}}$ and $b_{\mathrm{i}}$ fixed. In other words we
keep the coordinates $l_{\mathrm{i}}$ and $b_{\mathrm{i}}$ of each measurement and we randomly
shuffle the values $x_{\mathrm{i}}$ of each measurement. The actual values $x_{\mathrm{i}}$
remain unchanged, but they are randomly redistributed on the sky following the fixed measured
positions $l_{\mathrm{i}}$ and $b_{\mathrm{i}}$.

\item For each patch on the sky we calculate the test statistic $\xi^\mathrm{s}$ again, but this
time comparing the distributions of the shuffled data $G_\mathrm{q}^\mathrm{s}(x)$
for GRBs in each patch and the distribution for the whole sky $F_\mathrm{p}(x)$.
In what follows the superscript $s$, used with
any quantity, means that the quantity is related to randomly shuffled data.

\item We repeat the points 6. and 7. $n=100$ times. For some selected cases, when we want higher
precision of our results, we repeat points 6. and 7. $n=1\,000$ times (see further in the text).

\item For a given statistic $\xi$ we derive the limiting values $\xi^\mathrm{s}_{10}$,
$\xi^\mathrm{s}_{5}$, $\xi^\mathrm{s}_{1}$, and $\xi^\mathrm{s}_{0.1}$ which delimit the highest
10\,\%, 5\,\%, 1\,\%, and 0.1\,\% of all $\xi^\mathrm{s}$ values from all patches in all randomly
shuffled data, respectively. In what follows we denote these limiting values as
$\xi^\mathrm{s}_\mathrm{i}$, where i=10, 5, 1, or 0.1. $\xi^\mathrm{s}_\mathrm{i}$ is
derived from a set of $10^5$ or $10^6$ values in case we have $n=100$ or $n=1\,000$
shufflings (see Fig.~\ref{fig:method_4}).
For a given statistic $\xi$ we also derive value $\xi^\mathrm{s}_0$ which is the maximum of
all $\xi^\mathrm{s}$ values in all patches of all shuffled data. The vales $\xi^\mathrm{s}_0$ are
different for $n=100$ or $n=1\,000$ number of shufflings because they are derived from
a set of $10^5$ or $10^6$ values, respectively.

\item We compare the distributions of a given statistic $\xi$ for the measured data
and for all data shufflings. We count the number of patches $N^\mathrm{m}_\mathrm{i}$
in the measured data for which $\xi^\mathrm{m}>\xi^\mathrm{s}_\mathrm{i}$
(see Fig.~\ref{fig:method_5}).

\item The mean number of patches $\overline{N^\mathrm{s}_\mathrm{i}}$ in the randomly shuffled
data for which $\xi^\mathrm{s}>\xi^\mathrm{s}_\mathrm{i}$ is $\overline{N^\mathrm{s}_\mathrm{i}}$
= 100, 50, 10, and 1 for i = 10, 5, 1, and 0.1, respectively. This means that, for example,
if we find $N^\mathrm{m}_\mathrm{1} \gg 10$ it could indicate anisotropy in the measured data.

\item In the last step we calculate the probability (significance) $P^\mathrm{N}_\mathrm{i}$
of finding at least $N^\mathrm{m}_\mathrm{i}$ number of patches with
$\xi^\mathrm{s}>\xi^\mathrm{s}_\mathrm{i}$ in the randomly shuffled data
(see Fig.~\ref{fig:method_6}).

\item Perform all steps for several radii of the patches $r=20^\circ$, $30^\circ$, $40^\circ$,
$50^\circ$, $60^\circ$, for all tested quantities in our data sample $x=T_{90}$, $F_{64}$,
$F_{256}$, $F_{1024}$, $F_{64\mathrm{,B}}$, $F_{256\mathrm{,B}}$, $F_{1024\mathrm{,B}}$,
$S_{\mathrm{B}}$, $S$ and for all test statistics $\xi=AD$, $D$, $V$, and $\chi^2$.

\end{enumerate}

Concerning our method one may object that the tested samples should be statistically
independent, i.e. the sample of the whole sky should be independent of that being inside
a patch. However, as mentioned in point 12., we derive the significances
$P^\mathrm{N}_\mathrm{i}$ from the distribution of
$N^\mathrm{s}_\mathrm{i}(\xi^\mathrm{s}>\xi^\mathrm{s}_\mathrm{i})$
in the randomly shuffled data, that means from a true distribution describing a randomized
isotropic situation. We do not derive the significances $P^\mathrm{N}_\mathrm{i}$ from any
theoretical distributions. Thus in our method the independence of the tested samples is
not required.

We used routines ``KSTWO'' and ``KUIPERTWO'' of the
IDL\footnote{http://www.harrisgeospatial.com/ProductsandSolutions/\\
GeospatialProducts/IDL.aspx} Astronomy Users Library\footnote{http://idlastro.gsfc.nasa.gov/}
\citep{lan93} for calculation of the $D$ and $V$ statistics, respectively.
For the calculation of A--D statistic we employed the ``adk'' package
\footnote{https://cran.r-project.org/src/contrib/Archive/adk/} \citep{adk}
of the R software\footnote{https://www.r-project.org} \citep{rsoft}.

\begin{figure}[t]
\includegraphics[width=0.46\textwidth]{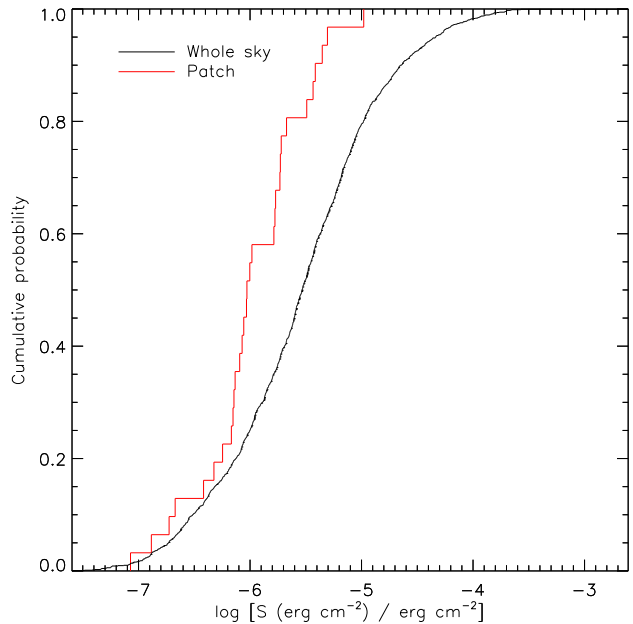}
\caption{An example of two empirical cumulative distribution functions $F_\mathrm{p}(x)$
and $G_\mathrm{q}(x)$ of a GRB property $x$ obtained for all GRBs in the whole sky (black curve)
and for all GRBs in one example sky patch (red curve), respectively. Both distributions are being
compared by several test statistics. In this case $x$ is fluence $S$.
This particular sky patch has radius of $r=20^\circ$ and center at the galactic coordinates
of $l=28.6^\circ$ and $b=16.9^\circ$. It is the patch for which
$D^\mathrm{m}_\mathrm{0}>D^\mathrm{s}_\mathrm{0}$ (see Table~\ref{tab:results_ks}).}
\label{fig:method_2}
\end{figure}

\section{Results}
\label{sec:results}

\begin{figure}[t]
\includegraphics[width=0.45\textwidth]{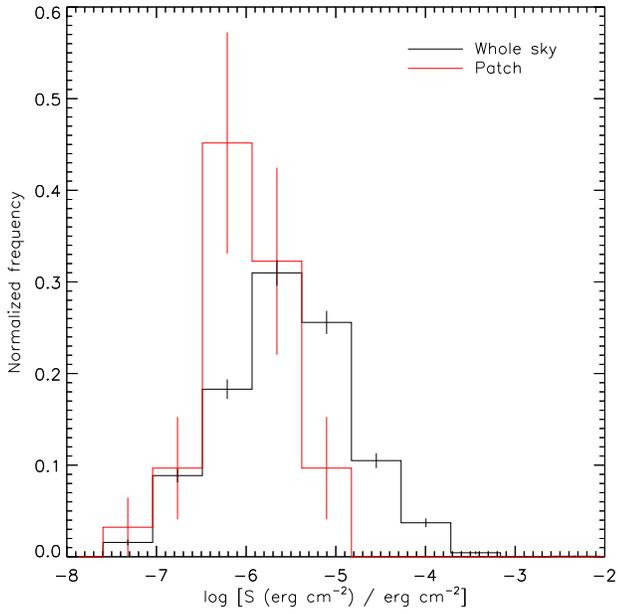}
\caption{An example of two distributions of a GRB property $x$
obtained for the whole sky and for one example sky patch,
which are being compared by $\chi^2$ statistic. In this case $x$ is fluence $S$.
The black curve denotes the normalized frequencies $Q_\mathrm{i}/Q$ for the whole sky,
where $Q_\mathrm{i}$ is the frequency in \textit{i}-th bin and $Q$=1591 is the
total number of events in the whole sky.
The red curve denotes the normalized frequencies $R_\mathrm{i}/R$ for a sky patch,
where $R_\mathrm{i}$ is the frequency in \textit{i}-th bin and $R$=31 is the
total number of events in the patch.
This particular sky patch is the same as in Fig.~\ref{fig:method_2} and it
has radius of $r=20^\circ$ and center at the galactic coordinates of
$l=28.6^\circ$ and $b=16.9^\circ$.
The error bars over each bin are Poisson errors normalized by $Q$ or $R$,
respectively.}
\label{fig:method_3}
\end{figure}

This section describes the results obtained by the proposed method applied on the
{\em Fermi} / GBM data sample. Fig.~\ref{fig:method_2} shows an example of two
empirical cumulative distribution functions of fluence $S$ obtained for the GRBs of the whole sky
and for GRBs of one example sky patch. Both distributions are being compared by $AD$, $D$,
and $V$ test statistics. Fig.~\ref{fig:method_3} shows the same distributions,
this time binned, which are being compared by $\chi^2$ statistic.

\begin{figure}[t]
\includegraphics[width=0.45\textwidth]{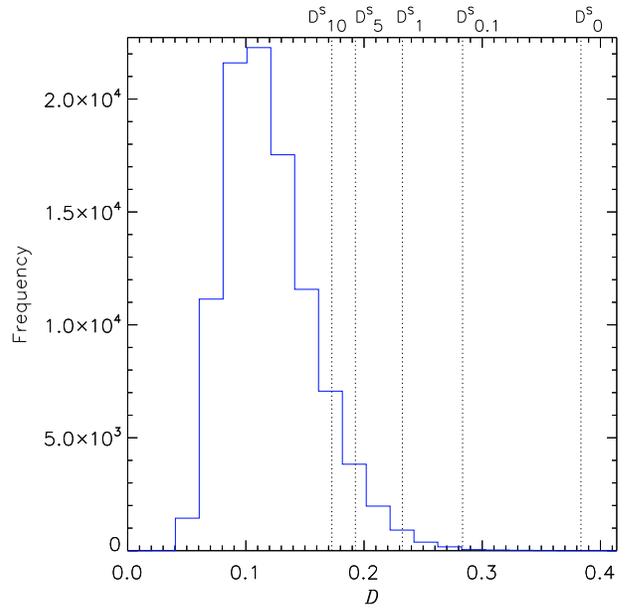}
\caption{An example of the distribution of the statistic $D$
obtained for all sky patches and all $n=100$ data shufflings.
Here the fluences $S$ and patch radii $r=20^\circ$ were used.
The values $D^\mathrm{s}_{10}$, $D^\mathrm{s}_{5}$, $D^\mathrm{s}_{1}$,
$D^\mathrm{s}_{0.1}$, and $D^\mathrm{s}_{0}$ are marked by
vertical dotted lines.}
\label{fig:method_4}
\end{figure}

Fig.~\ref{fig:method_4} shows an example of the distribution of the statistic $D$
obtained for all sky patches of radii $r=20^\circ$, all $n=100$ data shufflings, and
fluences $S$. From such a distribution $D^\mathrm{s}_{10}$, $D^\mathrm{s}_{5}$,
$D^\mathrm{s}_{1}$, $D^\mathrm{s}_{0.1}$, and $D^\mathrm{s}_{0}$ were calculated.

\begin{figure}[t]
\includegraphics[width=0.45\textwidth]{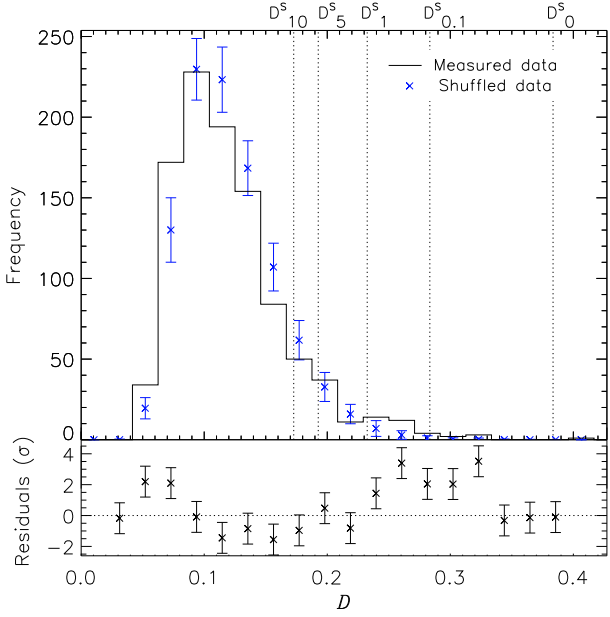}
\caption{An example of the comparison of the distributions of the statistic $D$
for the measured ($D^\mathrm{m}$) and shuffled ($D^\mathrm{s}$) data.
Here the fluences $S$ and patch radii
$r=20^\circ$ were used. The blue crosses denote the mean frequencies
for given bins and the error bars are corrected sample standard
deviations $\sigma$ for $n=100$ data shufflings.
The values $D^\mathrm{s}_{10}$, $D^\mathrm{s}_{5}$, $D^\mathrm{s}_{1}$,
$D^\mathrm{s}_{0.1}$, and $D^\mathrm{s}_{0}$ marked by vertical dotted lines
are the same as in Fig.~\ref{fig:method_4}.
For example, here the number of patches in the measured data for which
$D^\mathrm{m}>D^\mathrm{s}_\mathrm{1}$ is $N^\mathrm{m}_\mathrm{1}=34$
(see Table~\ref{tab:results_ks}).}
\label{fig:method_5}
\end{figure}

Fig.~\ref{fig:method_5} shows an example of the comparison of the distributions of
the statistic $D$ for the measured ($D^\mathrm{m}$) and shuffled ($D^\mathrm{s}$) data.

\begin{figure}[t]
\includegraphics[width=0.45\textwidth]{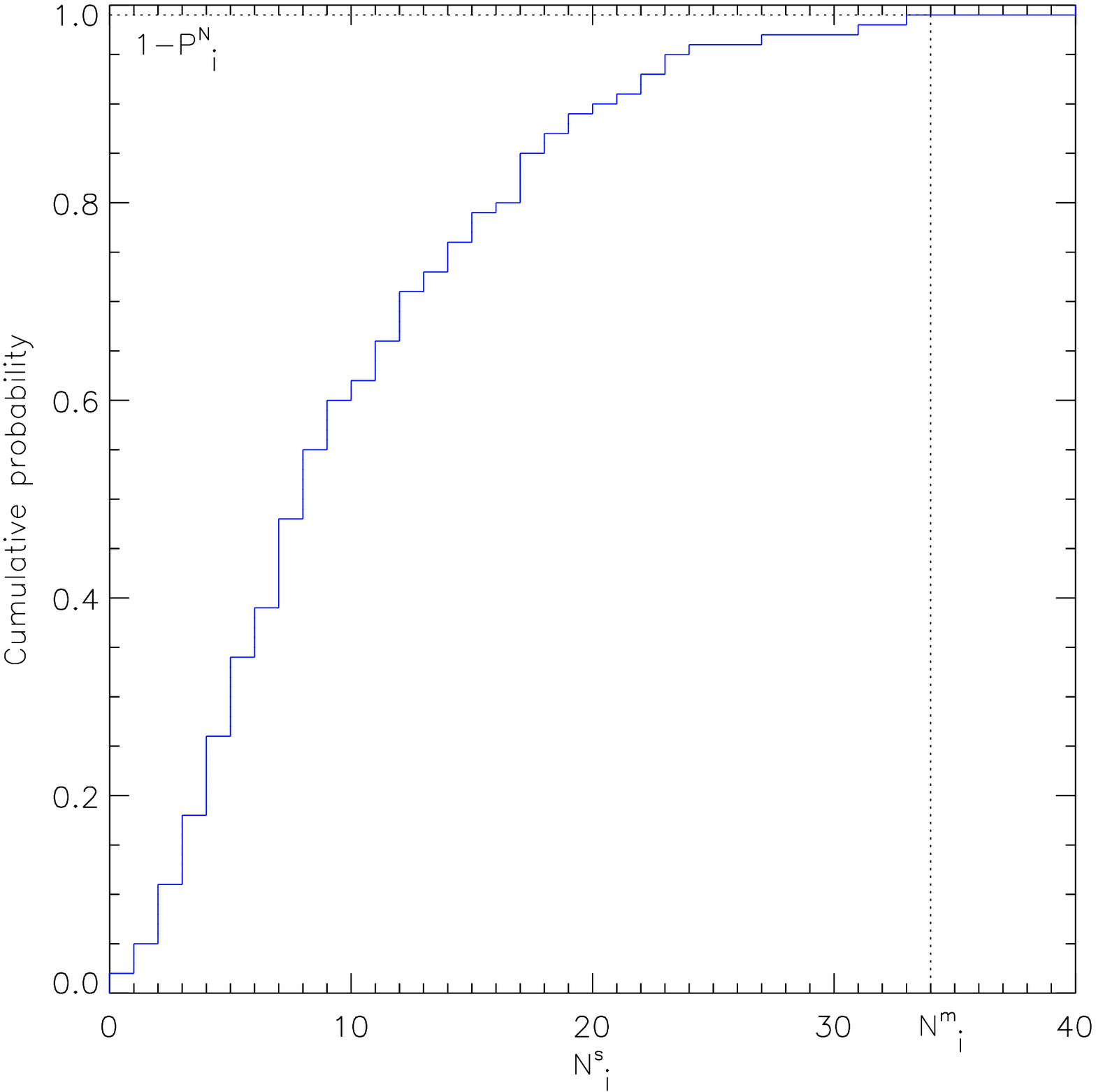}
\caption{An example of the cumulative distribution of $N^\mathrm{s}_\mathrm{i}$,
which is the number of patches with $\xi^\mathrm{s}>\xi^\mathrm{s}_\mathrm{i}$ in
the randomly shuffled data. $N^\mathrm{m}_\mathrm{i}$ is the number of patches
in the measured data for which $\xi^\mathrm{m}>\xi^\mathrm{s}_\mathrm{i}$.
The probability (significance) $P^\mathrm{N}_\mathrm{i}$ of finding at least
$N^\mathrm{m}_\mathrm{i}$ number of patches with $\xi^\mathrm{s}>\xi^\mathrm{s}_\mathrm{i}$
in the randomly shuffled data can be obtained from this distribution. 
This example result is for fluence $S$, $\xi=D$, $i=1$, patch radii $r=20^\circ$,
and $n=100$ data shufflings. Here $N^\mathrm{m}_\mathrm{i}=N^\mathrm{m}_\mathrm{1}=34$
and $P^\mathrm{N}_\mathrm{i}=P^\mathrm{N}_\mathrm{1}=1\,\%$ (see Table~\ref{tab:results_ks}).}
\label{fig:method_6}
\end{figure}

Fig.~\ref{fig:method_6} shows an example of the cumulative distribution of
$N^\mathrm{s}_\mathrm{1}$, which is the number of patches with
$D^\mathrm{s}>D^\mathrm{s}_\mathrm{1}$ in the randomly shuffled data.
This example result is for fluence $S$, patch radii $r=20^\circ$, and $n=100$ data
shufflings. If we have $N^\mathrm{m}_\mathrm{1}$ the number of patches
in the measured data for which $D^\mathrm{m}>D^\mathrm{s}_\mathrm{1}$, then from this
figure one can obtain the chance probability (significance) $P^\mathrm{N}_\mathrm{1}$
of finding at least $N^\mathrm{m}_\mathrm{1}$ number of patches with
$D^\mathrm{s}>D^\mathrm{s}_\mathrm{1}$ in the randomly shuffled data.

Table~\ref{tab:results_ks} of Appendix~\ref{sec:append_tables}
summarizes the results of all tested fluxes, fluences and duration
of GRBs in the data sample using the Kolmogorov--Smirnov statistic $D$. The most important are
columns $N^\mathrm{m}_\mathrm{i}$ and $P^\mathrm{N}_\mathrm{i}$. $N^\mathrm{m}_\mathrm{i}$ is
the number of patches with $D^\mathrm{m}>D^\mathrm{s}_\mathrm{i}$ in the randomly shuffled data,
where i=10, 5, 1, or 0.1. $P^\mathrm{N}_\mathrm{i}$ is the probability of finding at least
$N^\mathrm{m}_\mathrm{i}$ number of patches with $D^\mathrm{s}>D^\mathrm{s}_\mathrm{i}$
in the randomly shuffled data. The randomly shuffled data are randomized samples and thus
represent hypothetical isotropic samples. The cases with $P^\mathrm{N}_\mathrm{i}\leq 5$\,\%,
are emphasized in boldface.

Concerning the fluences $S$, $S_\mathrm{B}$, peak fluxes $F_{64}$, $F_{256}$, $F_{1024}$,
$F_{64\mathrm{,B}}$, $F_{256\mathrm{,B}}$ and patch radii $r=20^\circ$ or $r=30^\circ$
we obtained $P^\mathrm{N}_\mathrm{5}$, $P^\mathrm{N}_\mathrm{1}$, or $P^\mathrm{N}_\mathrm{0.1}$
$\leq 5$\,\%.

The most prominent discrepancy between the actual measured data and the randomly shuffled data
is found for the peak flux $F_{64}$, patch radii $r=20^\circ$ and $n=1000$ data shufflings. 
The mean number of patches $\overline{N^\mathrm{s}_\mathrm{0.1}}$ in the randomly shuffled
data for which $D^\mathrm{s}>D^\mathrm{s}_\mathrm{0.1}$ should be
$\overline{N^\mathrm{s}_\mathrm{0.1}}=1$ because we applied 1000 sky patches.
However, the actual measured data gives $N^\mathrm{m}_\mathrm{0.1}=14 \gg 1$.
The chance probability of finding at least 14 patches on the sky with
$D^\mathrm{s}>D^\mathrm{s}_\mathrm{0.1}$ in the randomly shuffled data is only 0.4\,\%.
This may indicate an anomaly in the measured data when compared to the
simulated isotropic data samples.

Moreover, there are five cases concerning the fluences $S$, $S_\mathrm{B}$, peak fluxes
$F_{64}$, $F_{256}$, and $F_{1024}$ (patch radii $r=20^\circ$ and 100 random data
shufflings), where a patch gives the value of $D^\mathrm{m}$ in the measured data
higher than the highest statistic $D^\mathrm{s}_{0}$ of all patches in all randomly shuffled data.

Table~\ref{tab:results_kuiper} of Appendix~\ref{sec:append_tables}
summarizes the results of all tested GRB
quantities in the data sample using the Kuiper statistic $V$.
Again, the cases with $P^\mathrm{N}_\mathrm{i}\leq 5$\,\%, are emphasized in boldface.

Concerning the fluence $S$, peak fluxes $F_{64}$, $F_{1024}$, $F_{64\mathrm{,B}}$
and patch radii $r=20^\circ$ or $r=40^\circ$ we obtained $P^\mathrm{N}_\mathrm{10}$,
$P^\mathrm{N}_\mathrm{5}$, $P^\mathrm{N}_\mathrm{1}$, or $P^\mathrm{N}_\mathrm{0.1}$
$\leq 5$\,\%.

The most prominent discrepancy between the actual measured data and the randomly shuffled data
is, similarly to the $D$ statistic, found for the peak flux $F_{64}$, patch radii $r=20^\circ$
and $n=1000$ data shufflings. The actual measured data gives
$N^\mathrm{m}_\mathrm{0.1}=7 \gg \overline{N^\mathrm{s}_\mathrm{0.1}} = 1$.
The chance probability of finding at least 7 patches on the sky with
$V^\mathrm{s}>V^\mathrm{s}_\mathrm{0.1}$ in the randomly shuffled data is only 1.9\,\%.

Moreover, there are three cases concerning the fluence $S$, peak fluxes $F_{256}$, and $F_{1024}$
(patch radii $r=20^\circ$ and 100 random data shufflings), where a patch gives the value of
$V^\mathrm{m}$ in the measured data higher than the highest statistic $V^\mathrm{s}_{0}$
of all patches in all randomly shuffled data.

Table~\ref{tab:results_ad} of Appendix~\ref{sec:append_tables}
summarizes the results of all tested GRB quantities in the data
sample using the Anderson--Darling statistic $AD$.
The cases with $P^\mathrm{N}_\mathrm{i}\leq 5$\,\%, are emphasized in boldface.

Concerning the fluence $S$, peak fluxes $F_{64}$, $F_{64\mathrm{,B}}$, $F_{256\mathrm{,B}}$
and patch radii $r=20^\circ$ or $r=30^\circ$ we obtained $P^\mathrm{N}_\mathrm{1}$, or
$P^\mathrm{N}_\mathrm{0.1}$ $\leq 5$\,\%.

The most prominent discrepancy between the actual measured data and the randomly shuffled data
is found for the peak flux $F_{256\mathrm{,B}}$, patch radii $r=20^\circ$
and $n=100$ data shufflings. The actual measured data gives
$N^\mathrm{m}_\mathrm{0.1}=10 \gg \overline{N^\mathrm{s}_\mathrm{0.1}} = 1$.
The chance probability of finding at least 10 patches on the sky with
$AD^\mathrm{s}>AD^\mathrm{s}_\mathrm{0.1}$ in the randomly shuffled data is only 1\,\%.
This may indicate an anomaly in the measured data when compared to the
simulated isotropic data samples.

\begin{figure*}
\begin{tabular}{cc}
\includegraphics[width=0.46\textwidth]{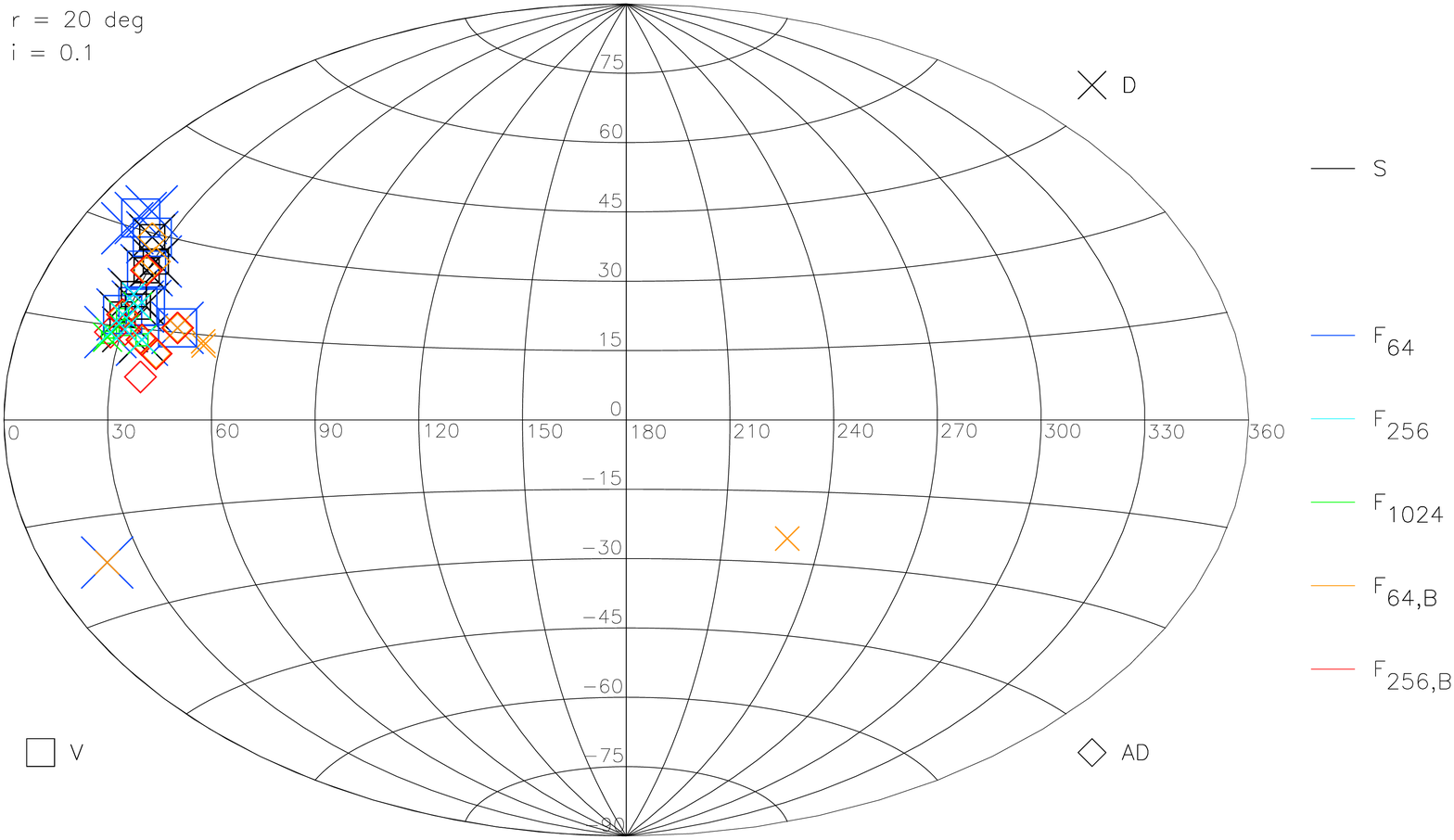} & \includegraphics[width=0.46\textwidth]{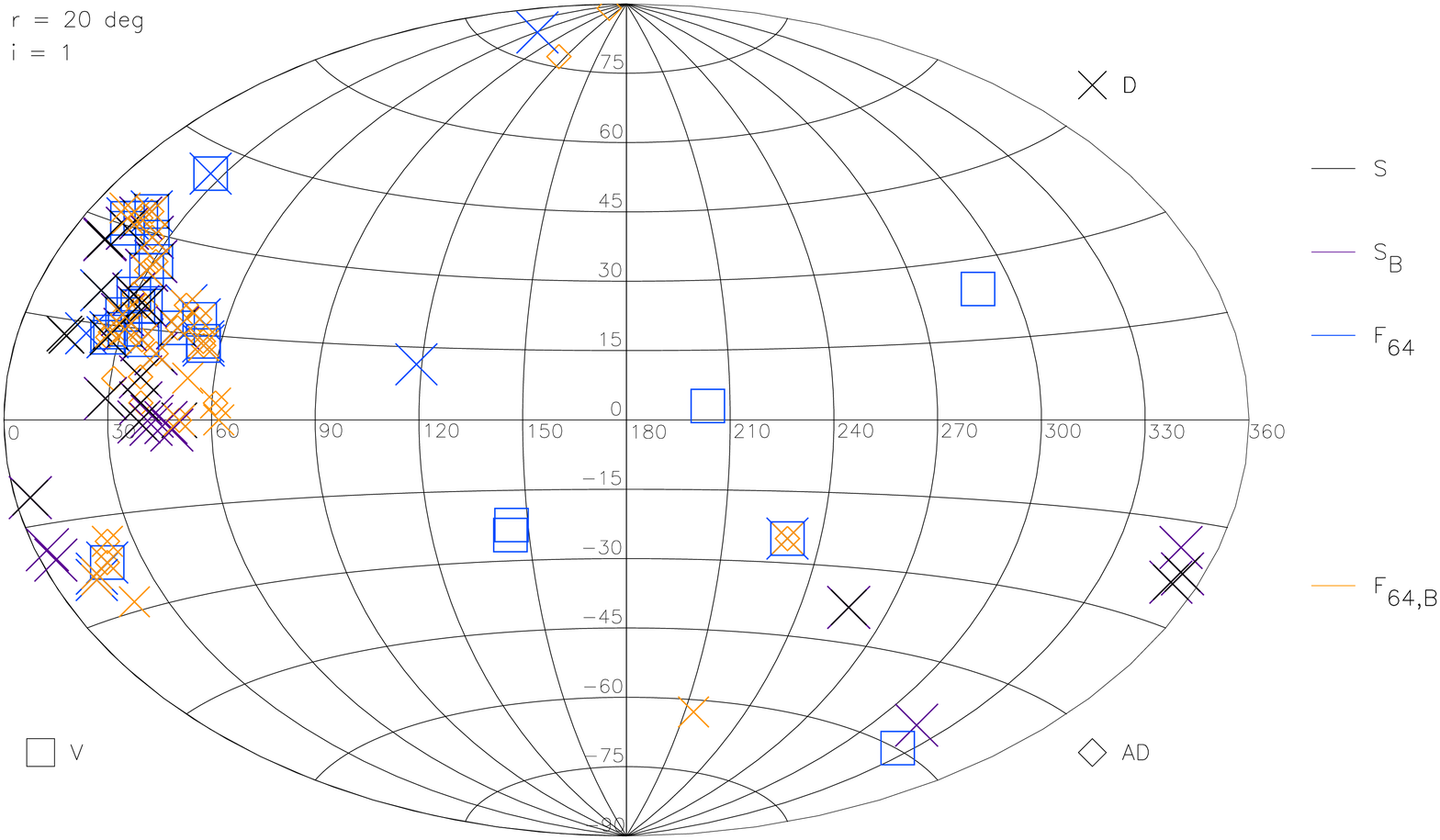} \\[1.0ex]
\includegraphics[width=0.46\textwidth]{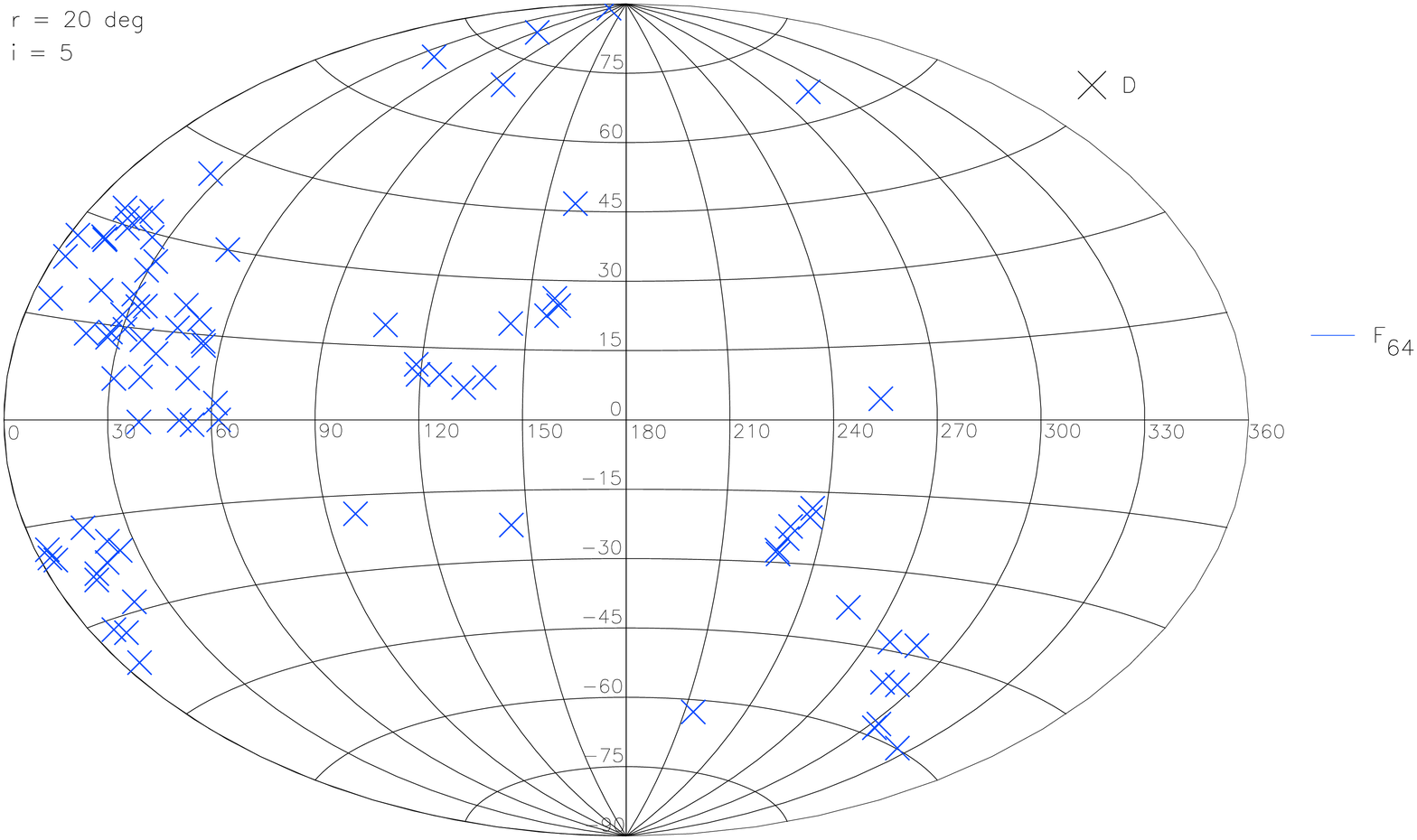} & \includegraphics[width=0.46\textwidth]{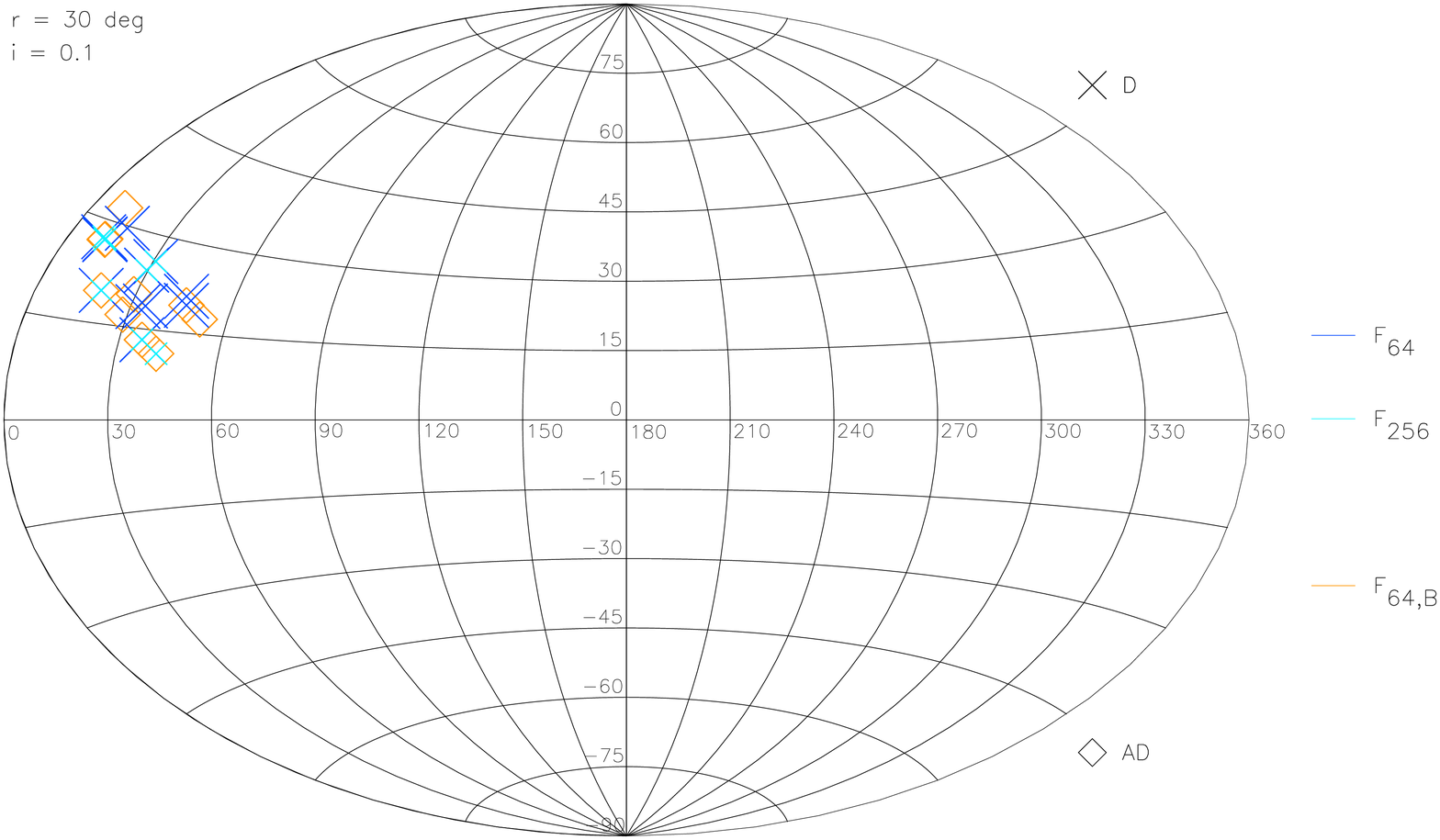} \\[1.0ex]
\includegraphics[width=0.46\textwidth]{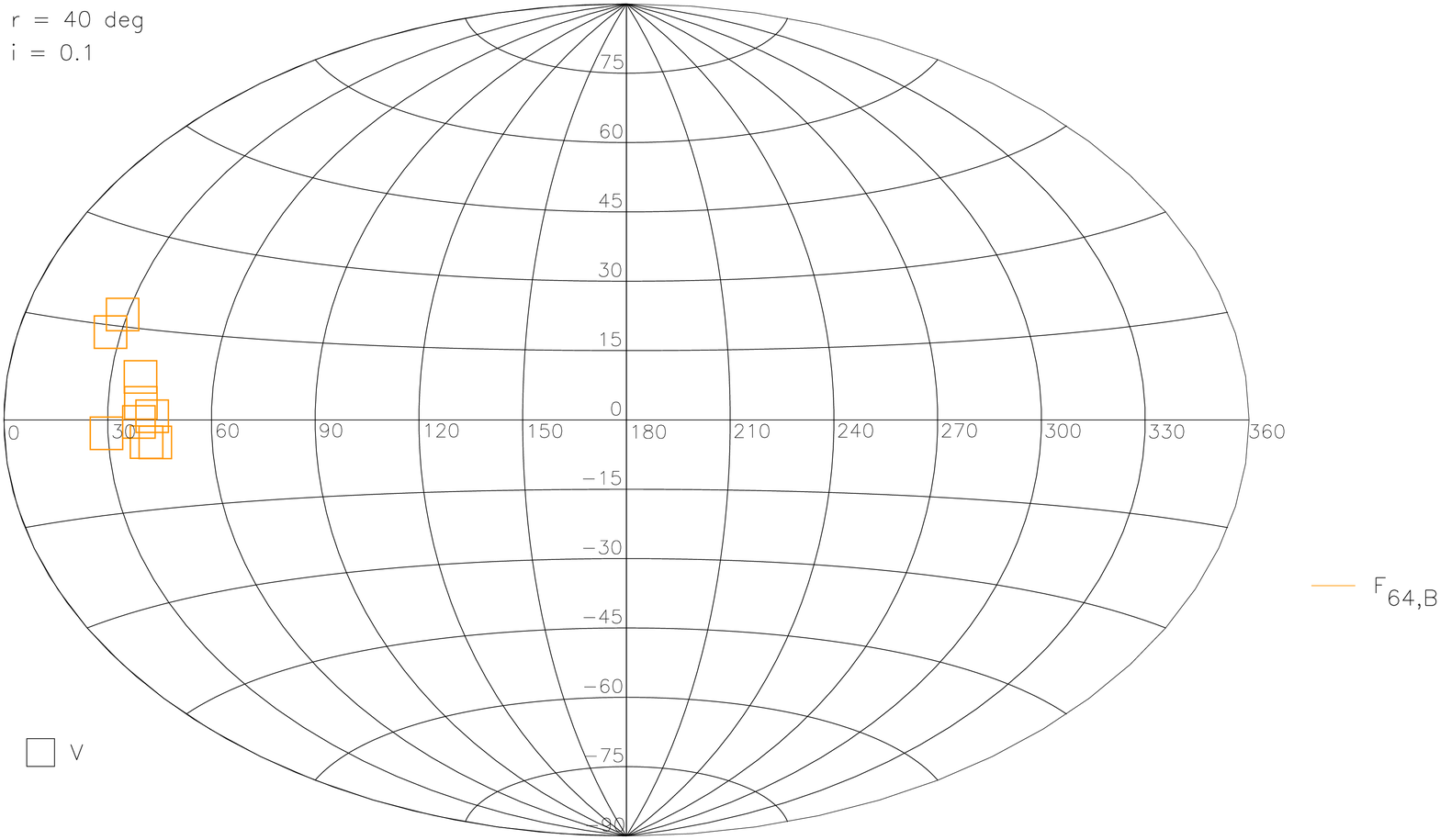} & \includegraphics[width=0.46\textwidth]{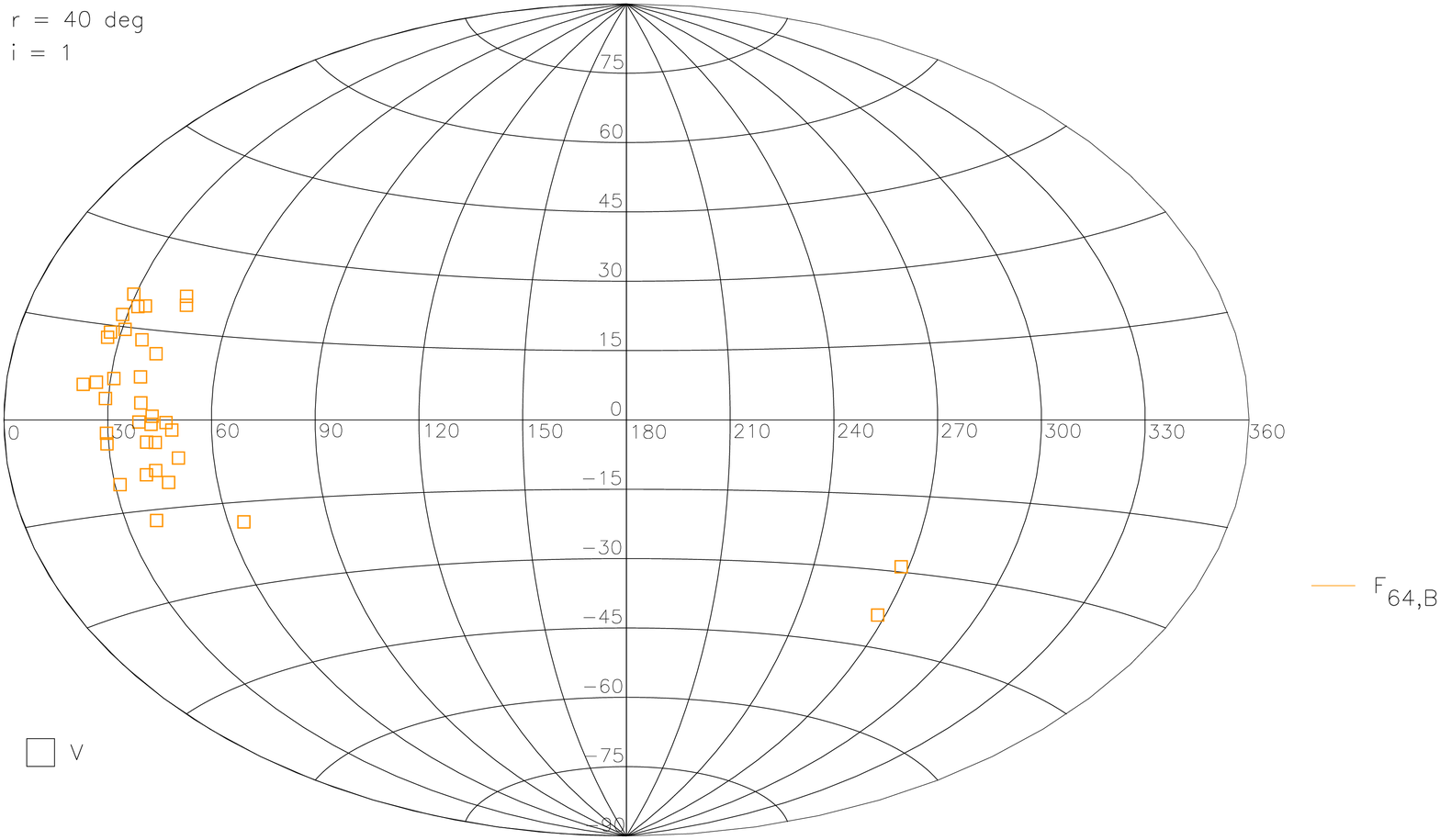} \\
\end{tabular}
\caption{Plotted are the patch centers on the sky in Galactic Coordinates (Aitoff projection),
for which the statistical properties of GRBs are mostly deviated from the randomness.
The combined results for different statistics $\xi=D, V$, or $AD$: Kolmogorov--Smirnov's $D$
(crosses), Kuiper's $V$ (squares), or Anderson--Darling's $AD$ (diamonds) are shown.
Specifically, the markers (crosses, squares, or diamonds) denote the centers of the patches
for which a given statistic $\xi^\mathrm{m}$, for the measured data, is higher than
$\xi^\mathrm{s}_\mathrm{i}$ obtained from the randomly shuffled data and the significance
$P^\mathrm{N}_\mathrm{i}\leq 5$\,\%, where i=5, 1, or 0.1.
The size of the markers is inverse proportional to $P^\mathrm{N}_\mathrm{i}$.
The exact values of $P^\mathrm{N}_\mathrm{i}$ can be found in
Tables~\ref{tab:results_ks} -- \ref{tab:results_chi2}.
Different colors mean different properties of GRBs being tested, particularly:
fluences $S$ (black), $S_{\mathrm{B}}$ (violet), and peak fluxes $F_{64}$ (blue), $F_{256}$
(light blue), $F_{1024}$ (green), $F_{64\mathrm{,B}}$ (orange), $F_{256\mathrm{,B}}$ (red).
In different pannels the results are plotted separately for different patch radii $r$
and for different $i$. The number of random data shufflings for these results is $n=1000$.}
\label{fig:results_1}
\end{figure*}

Moreover, there are two patches concerning the peak flux $F_{256\mathrm{,B}}$ (patch radius $r=20^\circ$
and 100 random data shufflings), which give the values of $AD^\mathrm{m}$ in the measured data
higher than the highest statistic $AD^\mathrm{s}_{0}$ of all patches in all randomly shuffled data.

Table~\ref{tab:results_chi2} of Appendix~\ref{sec:append_tables}
summarizes the results of all tested GRB quantities in the data
sample using the $\chi^2$ statistic. In the case of the $\chi^2$ statistic
the probabilities $P^\mathrm{N}_\mathrm{i}$ were never lower or equal 5\,\%.

Also there is no case of a patch, of a given radius and for a given tested quantity
in the measured data which gives $\chi^\mathrm{2\,m}$ higher than
the highest $\chi^\mathrm{2\,s}_{0}$ for patches of all random data shufflings.

Unlike the previous three tests the two-sample Chi-square test is applied on the binned
data of two samples instead of the unbinned empirical distribution functions. This may
be the reason why this test is not as sensitive as the other three tests.

Fig.~\ref{fig:results_1} shows the sky maps with marked patch centers for the
cases where $P^\mathrm{N}_\mathrm{i}\leq 5$\,\%
and for $\xi=D$, $V$, and $AD$ test statistics. It is interesting that for all three test statistics
and various fluxes and fluences there is a prominent area on the sky with the center
$l\approx 30^\circ$, $b\approx 15^\circ$ and radius $r\approx 20^\circ-40^\circ$,
where the most patches with $\xi^\mathrm{m}>\xi^\mathrm{s}_\mathrm{i}$ are concentrated.

Fig.~\ref{fig:results_2} shows the patch centers on the sky which give the values of
a given statistic $\xi^\mathrm{m}$ in the measured data higher than the highest statistic
$\xi^\mathrm{s}_{0}$ of all patches in all $n=100$ randomly shuffled data. Since we have
1000 patches on the sky and $n=100$, it means that those patches in the measured data give
the values of the statistic $\xi$ higher than any value obtained from the $10^5$ patches
in the randomly shuffled sample.

In Fig.~\ref{fig:method_4} and Fig.~\ref{fig:results_3}, we present the distributions
of the peak fluxes $F_{64}$, $F_{256}$, $F_{1024}$, $F_{64\mathrm{,B}}$, $F_{256\mathrm{,B}}$
and fluences $S$, $S_{\mathrm{B}}$ obtained for the whole sky and for the sky patch of
radius $r=20^\circ$ and its center at the galactic coordinates of $l=28.6^\circ$, $b=16.9^\circ$.
For this particular direction and size of the patch we obtained the values of $D^\mathrm{m}$
and $V^\mathrm{m}$ in the measured data higher than the highest statistics $D^\mathrm{s}_{0}$
and $V^\mathrm{s}_{0}$, respectively, of all patches in all $n=100$ randomly shuffled data.
It is the same direction as shown in Fig.~\ref{fig:results_2}.

\section{Discussion}
\label{sec:discuss}

\begin{figure*}[t]
\begin{tabular}{ccc}
\includegraphics[width=0.318\textwidth]{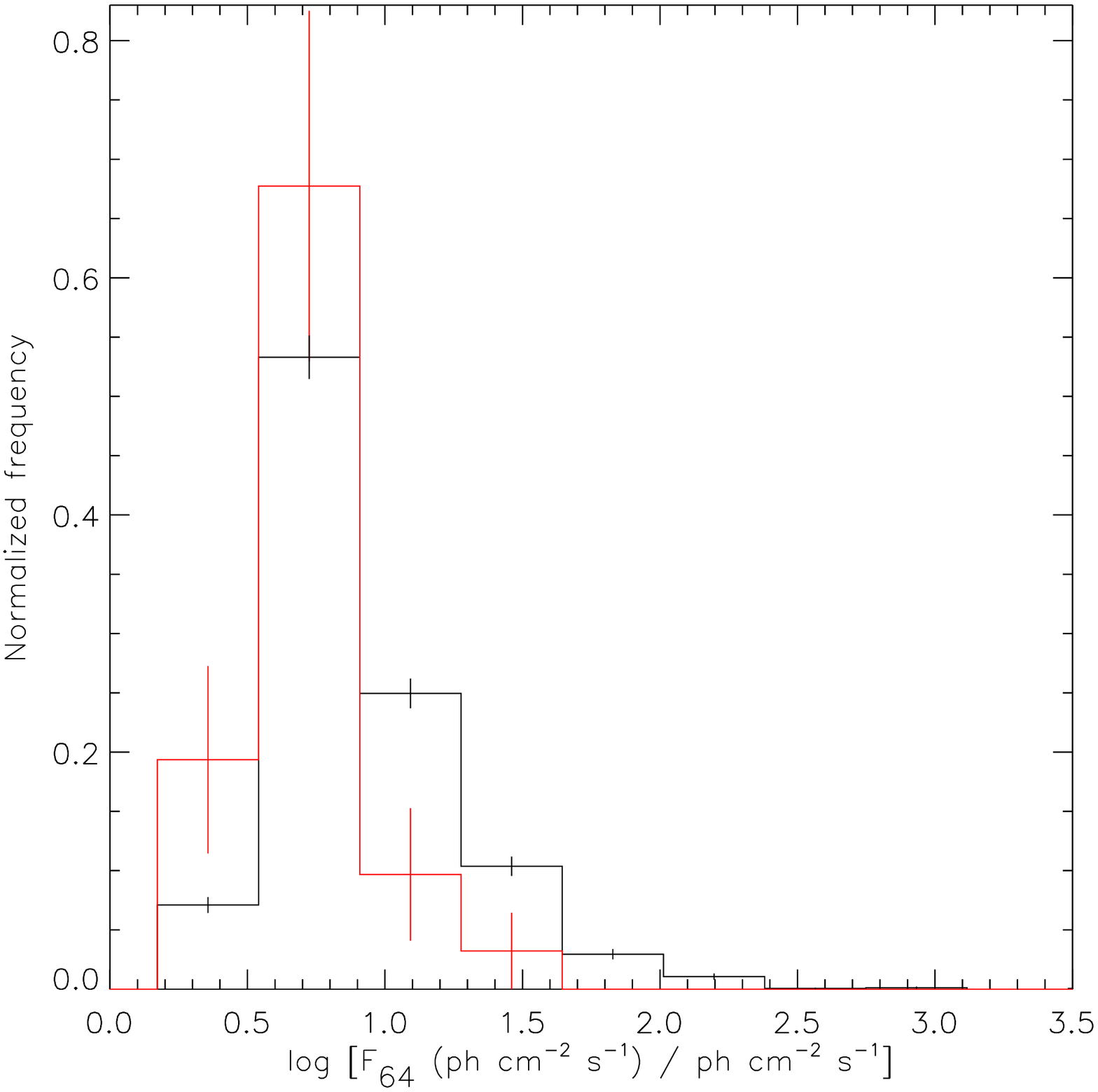} & \includegraphics[width=0.318\textwidth]{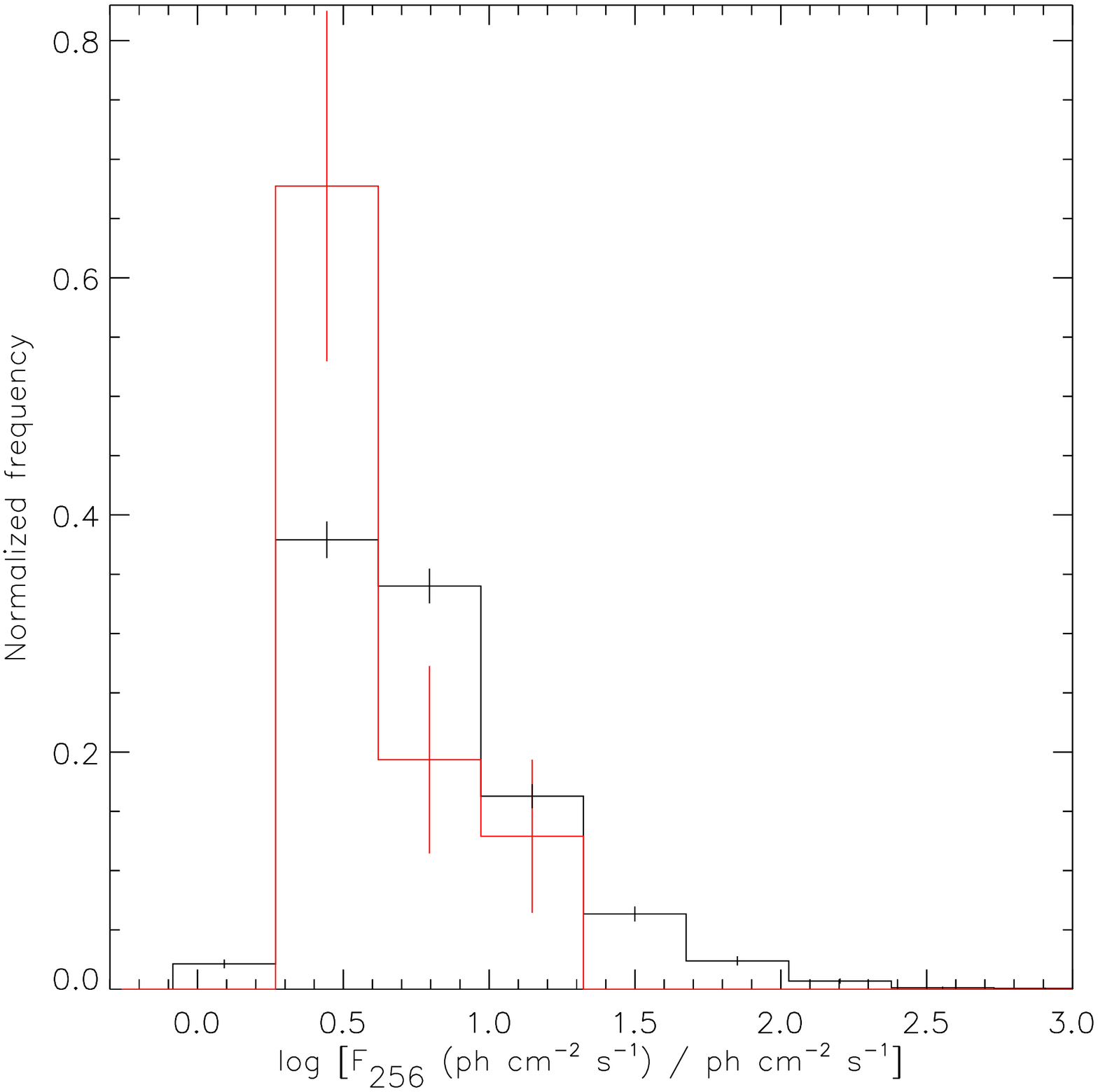} & \includegraphics[width=0.318\textwidth]{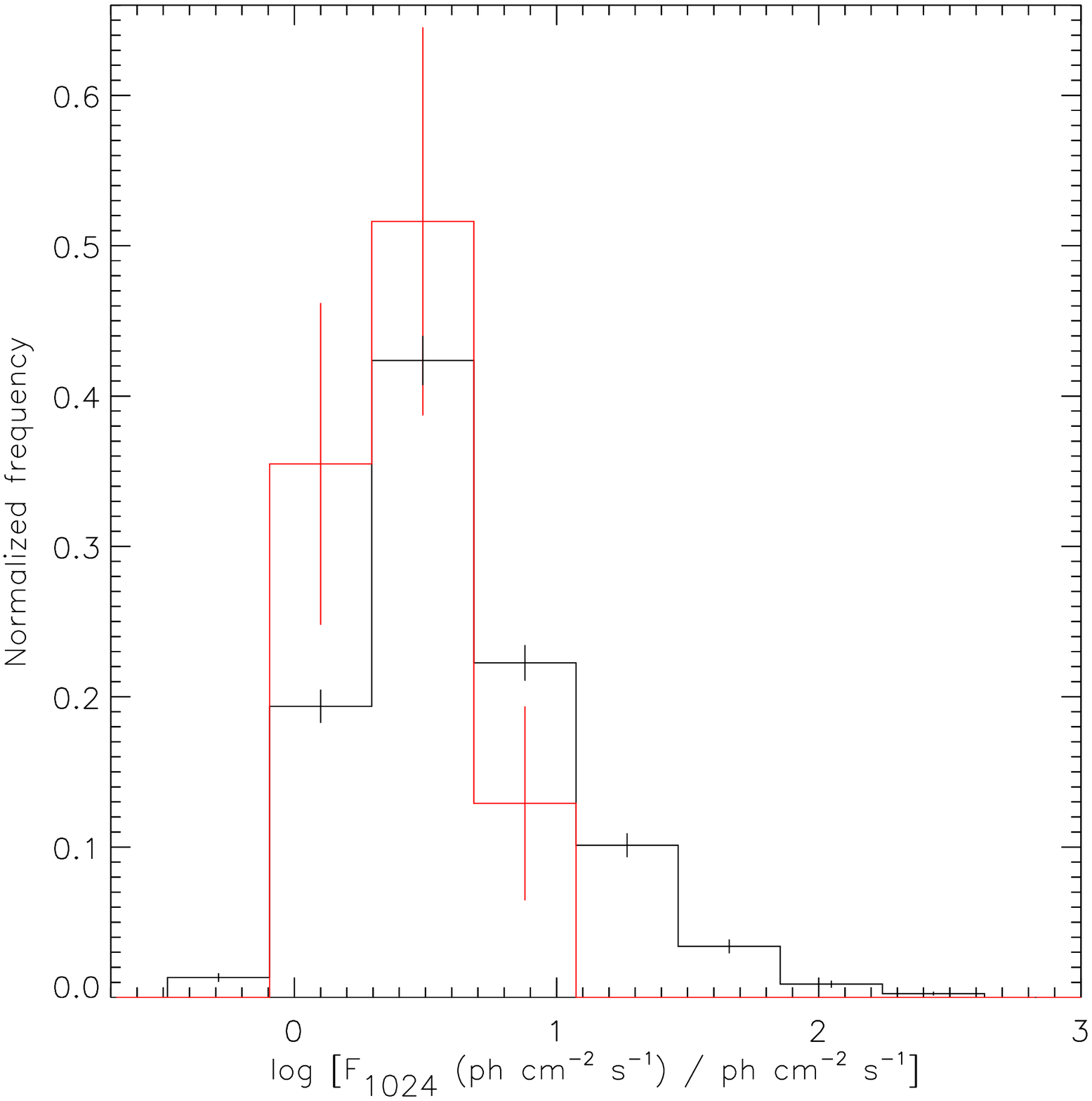} \\ [1.0ex]
\includegraphics[width=0.318\textwidth]{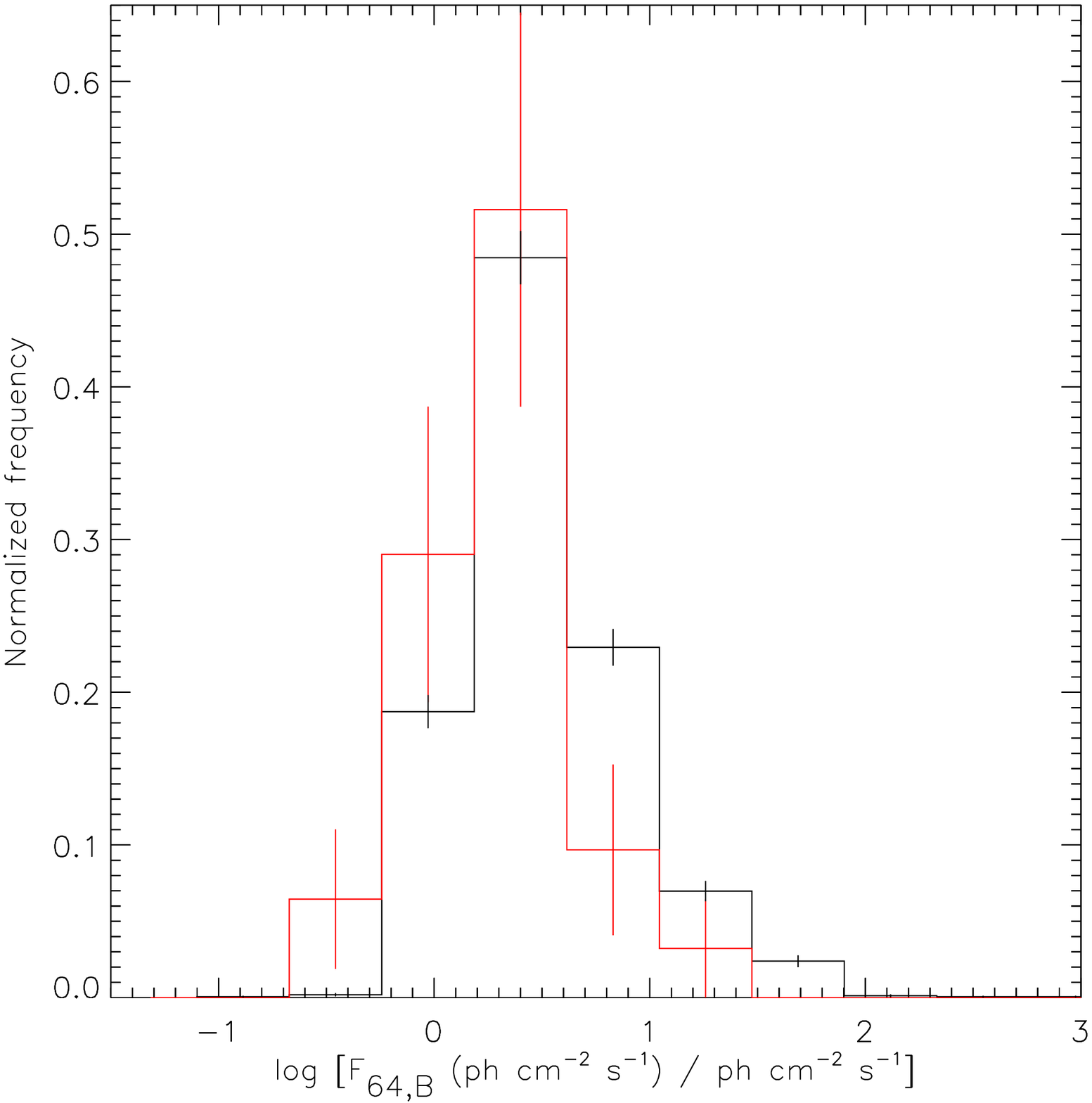} & \includegraphics[width=0.318\textwidth]{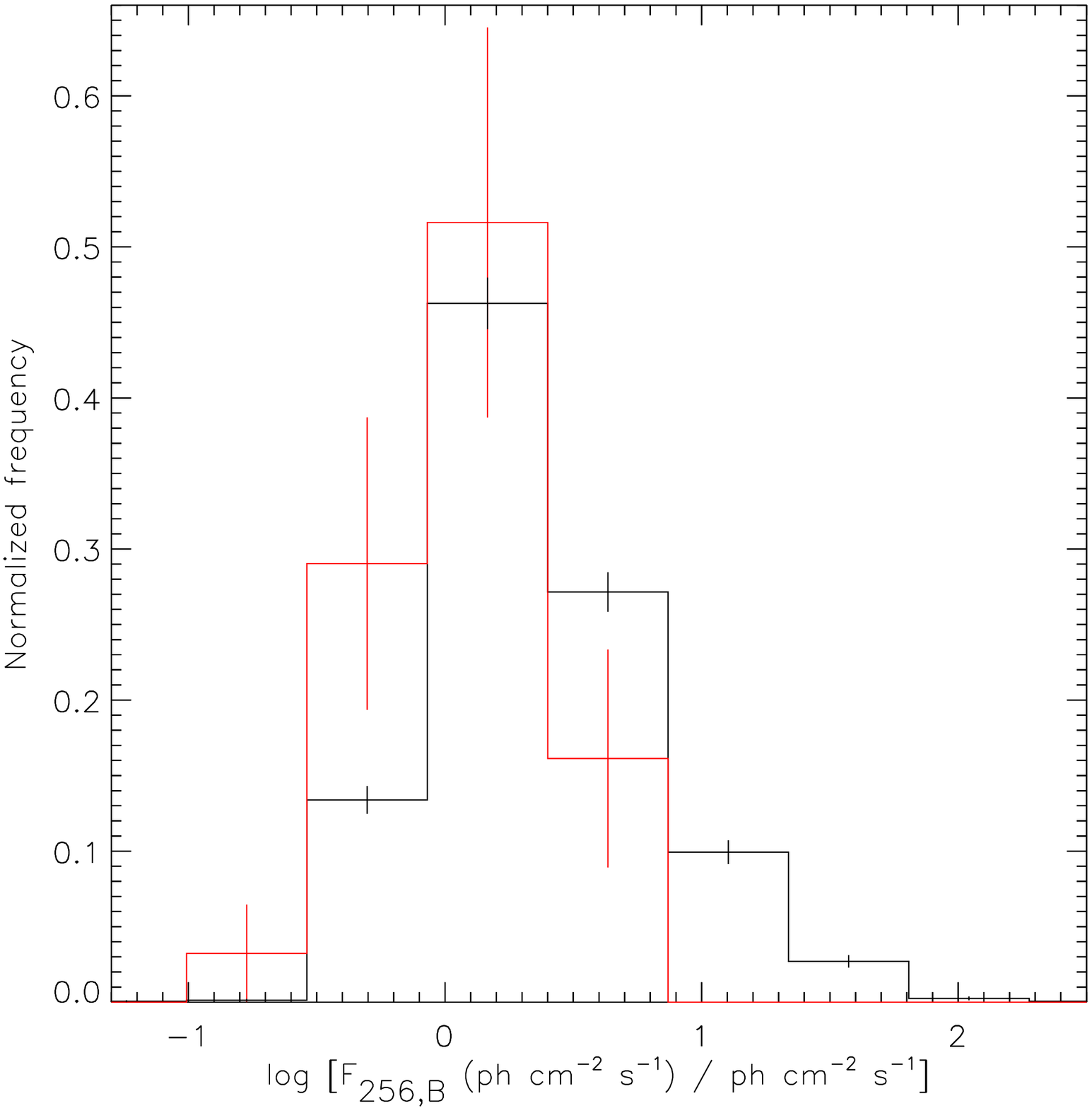} & \includegraphics[width=0.318\textwidth]{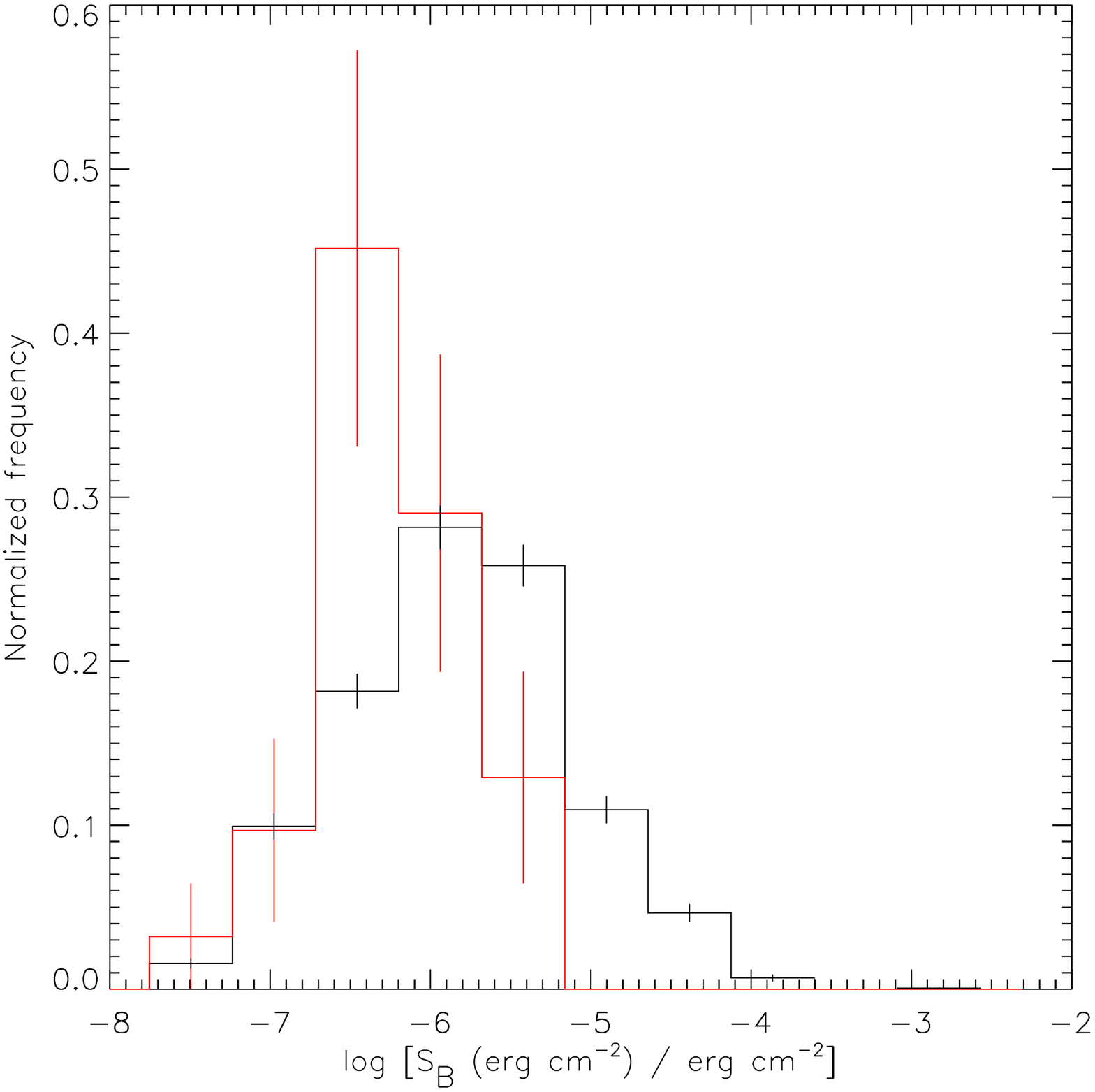} \\
\end{tabular}
\caption{Figure shows the distributions of the peak fluxes $F_{64}$, $F_{256}$,
$F_{1024}$, $F_{64\mathrm{,B}}$, $F_{256\mathrm{,B}}$ and fluence $S_{\mathrm{B}}$
obtained for the whole sky (black) and for one particular sky patch (red).
The sky patch has radius of $r=20^\circ$ and center at the galactic coordinates of
$l=28.6^\circ$, $b=16.9^\circ$.
The center of this patch is marked by black cross and square in Fig.~\ref{fig:results_2}.
The frequencies in each bin are normalized by the total number of events in the whole sky (1591)
and in the patch (31), respectively.
The error bars over each bin are Poisson errors, again normalized by the total number of events
in the whole sky and in the patch, respectively.}
\label{fig:results_3}
\end{figure*}

In order to check if our results are not caused by an artefact due to the initial
distribution of the patches on the sky, we again repeated our analysis starting with a
different set of 1\,000 randomly distributed patches on the sky and applied different
$n=100$ random data shufflings. We again obtained practically the same results.
There was a prominent area on the sky at the same position ($l\approx 30^\circ$,
$b\approx 15^\circ$ and radius $r\approx 20^\circ-50^\circ$) with the excess 
(compared to the randomly shuffled data) of the patches with the highest values of
the test statistics $D$, $V$, and $AD$. The probabilities $P^\mathrm{N}_\mathrm{i}\leq 5$\,\%
were obtained for fluxes $F_{64}$, $F_{256}$, $F_{1024}$, $F_{64\mathrm{,B}}$,
$F_{256\mathrm{,B}}$ and fluences $S_{\mathrm{B}}$, and $S$. The lowest chance probability
was $P^\mathrm{N}_\mathrm{i}<1$\,\%.

Let us discuss the overall significance of finding an anisotropic signal using all the
tests performed in this work. In our analysis we applied patches of five radii, nine different
GRB properties and four statistics. For one limiting value, e.g. $\xi^\mathrm{s}_{0.1}$, this suggests
180 statistical tests, however many of these tests are not independent of each other.
If we assume that there are 180 independent tests (trials) one can calculate the probability
($p-value$) of finding at least $k$ successes (each with probability $p$) from of the binomial test.
For details see for example Eq. (1) and (2) of \citet{vav08}.
From Tables~\ref{tab:results_ks} -- \ref{tab:results_chi2} one can see that
we obtained: $k=14$ for $p \leq 5$\,\% then $p-value=8.8$\,\%; $k=9$ for $p \leq 3$\,\% then $p-value=12.1$\,\%; $k=4$ for $p \leq 2$\,\% then $p-value=78.6$\,\%.
This is a course estimation because the tests are not independent.
The number of independent tests (trials) is likely tens.
We obtained that in one test the lowest chance probability obtained was 0.4\,\% (Kuiper's $V$
statistic, peak flux $F_{64}$, energy range $(10-1000)$\,keV and 64-ms timescale).
However, for example, if the number of independent tests is 10 then for $p=0.4$\,\% the
$p-value$ of the binomial test is 3.9\,\%. If the number of independent tests is 20
then the $p-value=7.7$\,\%. This suggests that a chance probability in a single test of 0.4\,\%
is actually much more likely to occur by chance in one of the many tests used.
Therefore likely our results do not point to a significant anisotropy.

It should be noted that the number of GRBs, in the patches centered at $l=28.6^\circ$ and
$b=16.9^\circ$, are 31, 97, and 160 for the patch radii $r=20^\circ, 30^\circ, 40^\circ$, respectively.
These numbers are lower than the mean values for all patches which are 48, 106, and 186 for the
same patch radii, respectively. From the distributions of the numbers of GRBs contained in all
patches of radii $r=20^\circ, 30^\circ, 40^\circ$ it follows that there are fractions of
99.0\,\%, 80.9\,\%, 97.5\,\%, respectively, of patches which contain the number of GRBs
higher than the number of GRBs in the patches centered at $l=28.6^\circ$ and $b=16.9^\circ$
for the same patch sizes. It means that the GRBs found in the anomalous patch centered at
$l=28.6^\circ$ and $b=16.9^\circ$ with radius $r=20^\circ$ do not only exhibit lower fluxes
and fluence, as shown in Fig~\ref{fig:method_3} and Fig~\ref{fig:results_3}, but also exhibit
lower number density than what the mean value is.
This is also demonstrated in Fig~\ref{fig:discuss} where the centers of patches
which represent the fraction of 5\,\% lowest occupied ones are plotted. The patch radii
$20^\circ$, $30^\circ$, and $40^\circ$ are plotted separately. The area, where we found
an anomaly, correlates with the less populated area on the sky.
One can expect that the area of reduced GRB density will have relatively
larger fluctuations in the measured GRB properties due to the Poisson noise.
The Poisson noise in the areas of reduced GRB density not only effects the measured data,
but also effects the randomly shuffled data. In our method we compare the number of patches
which have values of a statistic higher than a given limit in the measured data with the
number of patches having the same statistic higher than the same limit in the randomly
shuffled data. Therefore this comparison of the measured data with the shuffled ones should
make this method relatively robust.
When we increased the number of the simulated randomly shuffled data samples
from $n=100$ to $n=1000$ the statistical significance decreased.

\begin{figure}[t]
\includegraphics[width=0.45\textwidth]{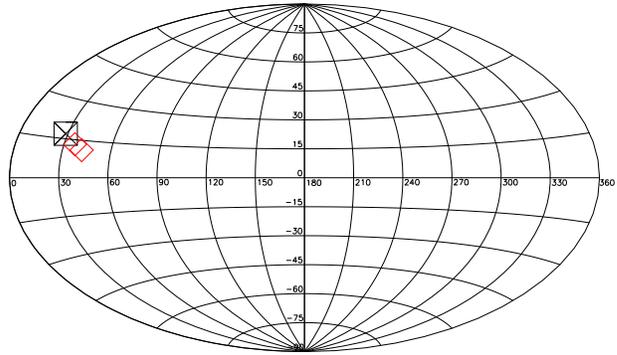}
\caption{Plotted are the patch centers on the sky in Galactic Coordinates
(Aitoff projection), which give the values of a given statistic $\xi^\mathrm{m}$
in the measured data higher than the highest statistic $\xi^\mathrm{s}_{0}$ of all
patches in $n=100$ randomly shuffled data. The red diamonds mark the patch centers
for Anderson--Darling's $AD$ statistic and peak flux $F_{256\mathrm{,B}}$.
The black cross marks one patch center for Kolmogorov--Smirnov's $D$
statistic and for five tested quantities: fluxes $F_{64}$, $F_{256}$, $F_{1024}$
and fluences $S$, $S_{\mathrm{B}}$.
The black square marks one patch center for Kuiper's $V$ statistic
and for three tested quantities: fluxes $F_{256}$, $F_{1024}$ and fluence $S$.
All cases are for patch radii $r=20^\circ$.}
\label{fig:results_2}
\end{figure}

\begin{figure}[t]
\begin{tabular}{c}
\includegraphics[width=0.45\textwidth]{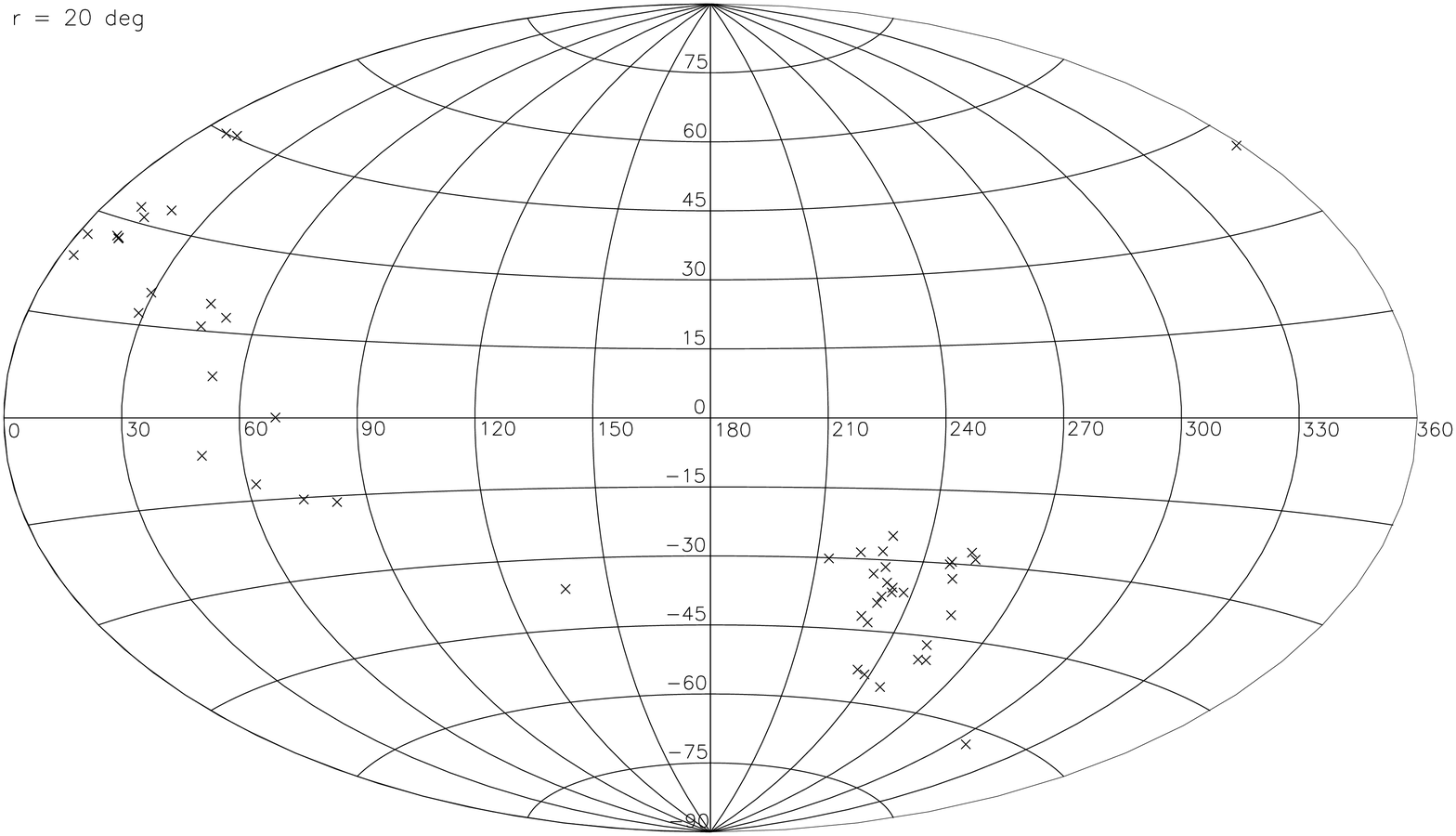} \\ [1.0ex]
\includegraphics[width=0.45\textwidth]{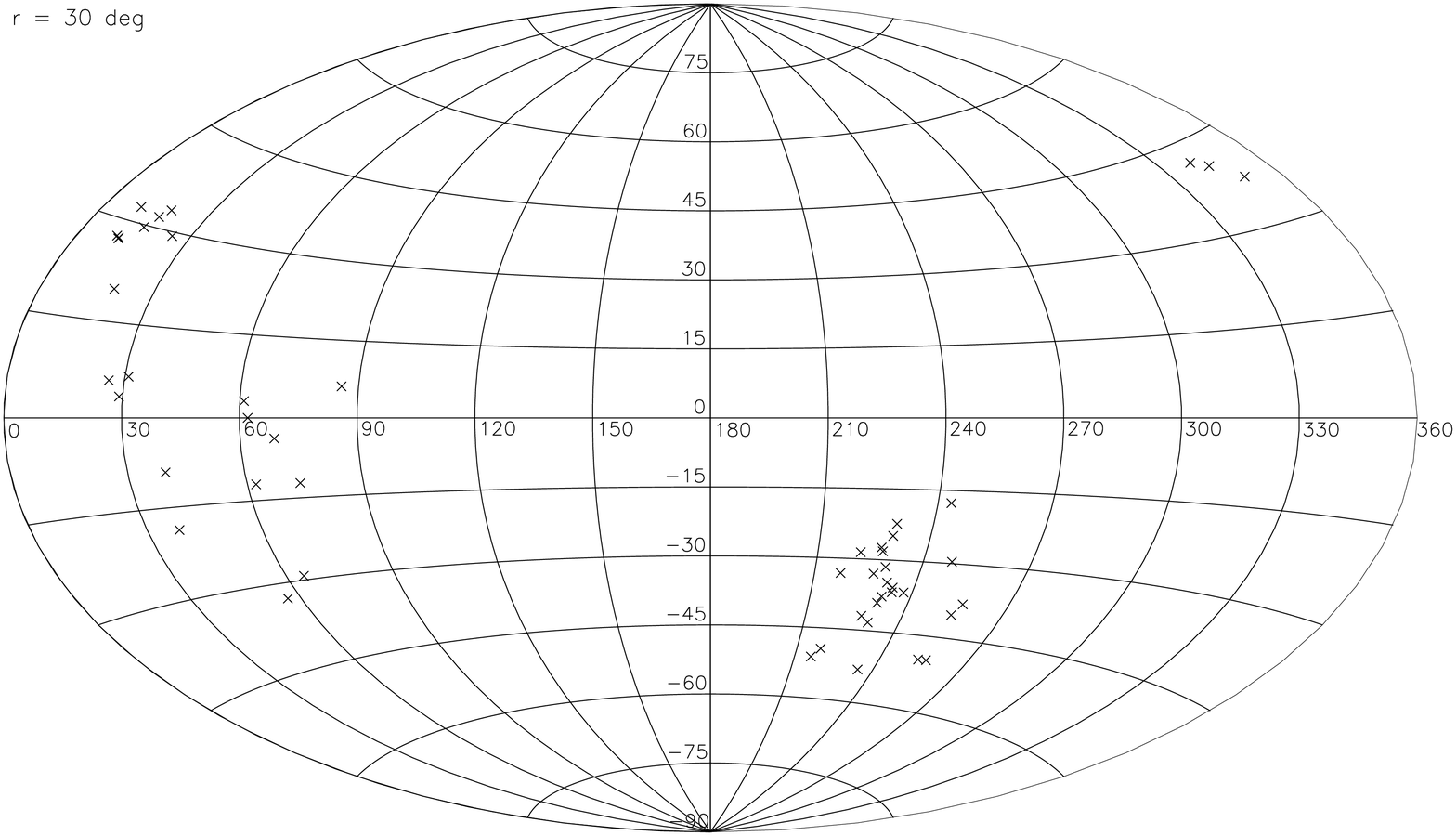} \\ [1.0ex]
\includegraphics[width=0.45\textwidth]{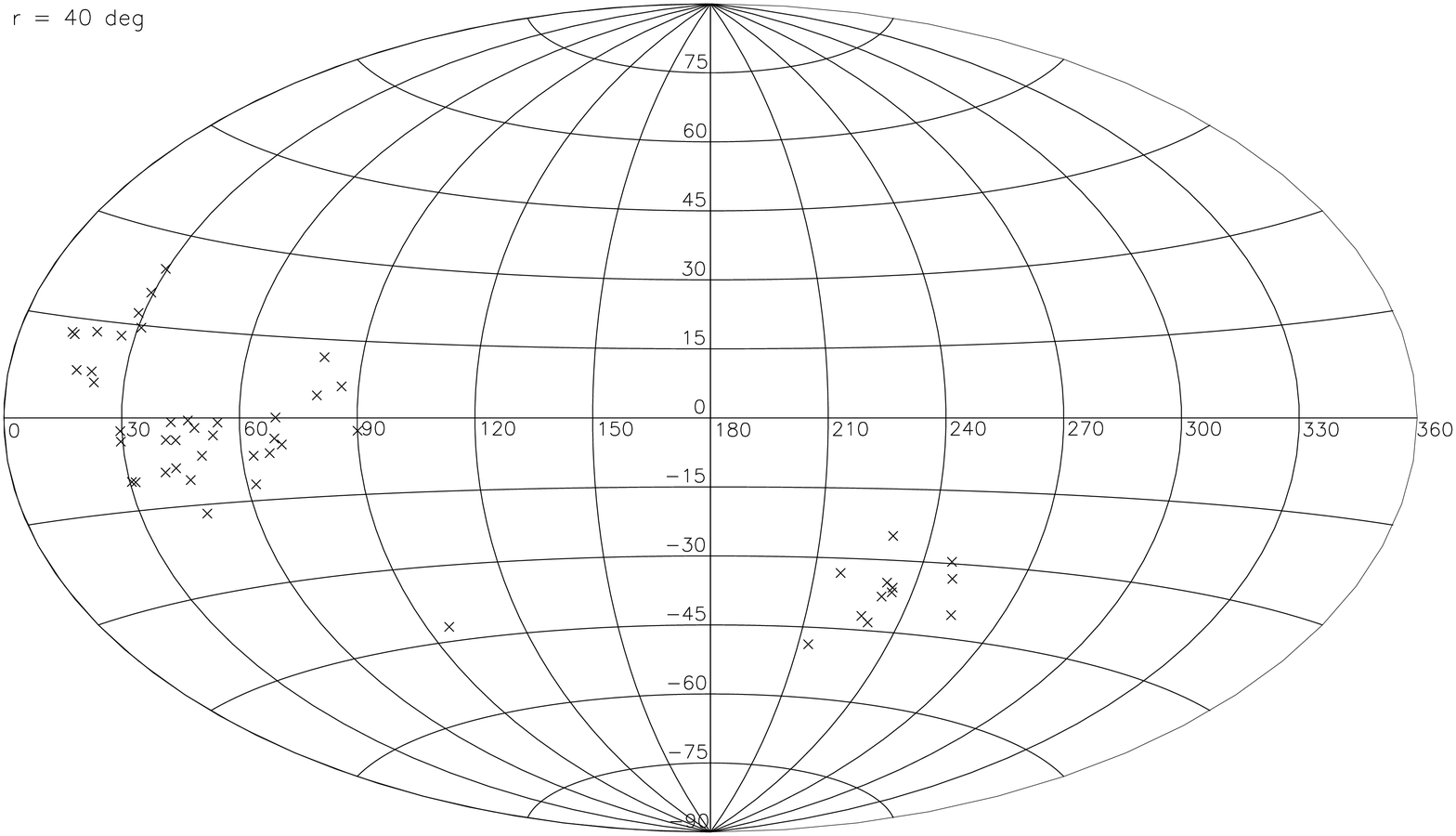} \\
\end{tabular}
\caption{Plotted are centers of the 5\,\% fraction of the lowest occupied patches
for patch radii $20^\circ$, $30^\circ$, and $40^\circ$.
}
\label{fig:discuss}
\end{figure}

One may suggest that the anomalous direction found in the {\em Fermi}/GBM data could be
due to some instrumental effects.
Areas on the sky of decreased exposure and lower density of observed GRBs are expected
for {\em Fermi}/GBM. The main reason is that some directions are less frequently observed due to
occultation by the Earth. The direction towards the Earth is not entirely randomized on the sky
because of the orbit and typical attitude of the {\em Fermi} satellite.
Therefore one can expect regions of systematically deficient exposure time and thus deficient
density of observed GRBs. Such effect was significant for Compton Gamma Ray
Observatory ({\em CGRO}) \citep{fis93,fis94,geh93,bri96,hak03}, which had a similar low Earth orbit to
{\em Fermi}. {\em CGRO} had inclination of $28.5^\circ$ and {\em Fermi} has inclination of
$25.6^\circ$.
Another question is if the Galactical gas and dust absorption can affect our results.

\citet{ho14,ho15} studied the spatial distribution of GRBs, detected by {\em Swift}
satellite \citep{geh04}, taking into account their redshift $z$. They used a two-dimensional
Kolmogorov-Smirnov test, a nearest-neighbour test, and a bootstrap point-radius method
applied on several sub-samples of different redshift intervals. They concluded that
there was a statistically significant clustering of the GRB sample at $1.6 < z \leq 2.1$
(Hercules--Corona Borealis Great Wall).
From their Figure~3 of \citet{ho14} or Figure~4 of \citet{ho15} it is seen that the
clustering happens at $l\approx 70^\circ$, $b\approx 45^\circ$ with a radius on the
sky roughly about $r\approx 40^\circ$. It is interesting that
the potential anomaly we found occurred
in somewhat similar direction, however not exactly in the same one.
Both areas are in the same quadrant on the sky
and they partially overlap.
This finding is also interesting considering that \citet{ho14,ho15}
used completely different approach and they studied number densities of GRBs on the sky
for various redshift intervals employing a data set from a different satellite.

\section{Conclusions}
\label{sec:conclude}

We tested the isotropy of the Universe through inspecting the isotropy of the properties
of GRBs such as their duration, fluences and peak fluxes at various energy bands and
different time scales. Summing up the results we conclude:
\begin{itemize}
\item We proposed a new method to test the isotropy of the Universe by testing the
observed properties of GRBs from large datasets.
The method was based on the comparisons of the distributions of a given measured GRB
property for a large number of randomly spread patches on the sky with a distribution of the
same GRB property for the whole sky. Four test statistics were applied to measure the
differences between the distributions of the GRB properties for random sky patches and for
the whole sky. The next key step was to compare the obtained distributions of the test
statistics derived from the measured data with the distributions of the test statistics
derived from randomly shuffled data to infer the significance of a potential anisotropy.

\item We applied the method on the {\em Fermi} / GBM data sample with 1591 GRBs.
We found that for three test statistics (Kolmogorov--Smirnov's $D$, Kuiper's $V$,
Anderson--Darling's $AD$) and various fluxes and fluences there is an area
on the sky with the center $l\approx 30^\circ$, $b\approx 15^\circ$ and radius
$r\approx 20^\circ-40^\circ$, where the most patches with the highest values of the test
statistics are concentrated. The inferred chance probabilities of observing the obtained excess
(compared to the randomly shuffled data) of the sky patches with high values of the test
statistics went down below 5\,\% depending on the tested quantity and the test statistic used.
However, many tests gave results consistent with isotropy.

\item While we noticed a feature in the {\em Fermi} / GBM data,
when we looked at the details more carefully we noticed a low number of GRBs in that
particular patch and when we increased the number of the simulated randomly shuffled
data samples the statistical significance decreased.
Likely our results do not point to a significant anisotropy.
Further investigation is highly desirable in order to confirm or reject these conclusions.
A larger {\em Fermi} / GBM data sample as well as data samples of other
GRB missions should be employed in the future and one should look for possible systematics.
\end{itemize}

\acknowledgments
We acknowledge use of the {\em Fermi} / GBM data.
This research has made use of data, software and/or web tools obtained from the High Energy
Astrophysics Science Archive Research Center (HEASARC), a service of the Astrophysics Science
Division at NASA/GSFC and of the Smithsonian Astrophysical Observatory's High Energy Astrophysics
Division. We kindly thank to L. G. Bal{\'a}zs, E. Linder, A. M{\'e}sz{\'a}ros, M. Tarnopolski,
P. Veres and an anonymous referee for useful comments and suggestions.
J.R. acknowledges the support of Taiwan's Ministry of Science and Technology (MOST)
funding number 105-2811-M-002-129. A.S. would like to acknowledge the support of the National
Research Foundation of Korea (NRF-2016R1C1B2016478).

\software{KSTWO.PRO and KUIPERTWO.PRO \citep{lan93}, adk: Anderson-Darling K-Sample Test and Combinations of Such Tests
\citep{adk}, R software \citep{rsoft}.}

\appendix

\section{Tables with results}
\label{sec:append_tables}

\begin{deluxetable*}{cccccccccccccccccc}
\tablecaption{\label{tab:results_ks}
Results using the Kolmogorov--Smirnov statistic $D$.}
\tablewidth{0pt}
\tabletypesize{\scriptsize}
\tablehead{
\colhead{Tested}	&	\colhead{$r$\tablenotemark{$\ast$}}	&	\colhead{$n$\tablenotemark{$\star$}}	&	\colhead{$D^\mathrm{s}_{10}$\tablenotemark{$\dagger$}}	&	\colhead{$N^\mathrm{m}_{10}$\tablenotemark{$\ddagger$}}	&	\colhead{$P^\mathrm{N}_{10}$\tablenotemark{$\#$}}	&	\colhead{$D^\mathrm{s}_{5}$\tablenotemark{$\dagger$}}	&	\colhead{$N^\mathrm{m}_{5}$\tablenotemark{$\ddagger$}}	&	\colhead{$P^\mathrm{N}_{5}$\tablenotemark{$\#$}}	&	\colhead{$D^\mathrm{s}_{1}$\tablenotemark{$\dagger$}}	&	\colhead{$N^\mathrm{m}_{1}$\tablenotemark{$\ddagger$}}	&	\colhead{$P^\mathrm{N}_{1}$\tablenotemark{$\#$}}	&	\colhead{$D^\mathrm{s}_{0.1}$\tablenotemark{$\dagger$}}	&	\colhead{$N^\mathrm{m}_{0.1}$\tablenotemark{$\ddagger$}}	&	\colhead{$P^\mathrm{N}_{0.1}$\tablenotemark{$\#$}}	&
\colhead{$D^\mathrm{s}_0$\tablenotemark{$\dagger$}}	&	\colhead{$N^\mathrm{m}_0$\tablenotemark{$\ddagger$}}	&	\colhead{$D^\mathrm{m}_0$\tablenotemark{\S}}
\\[-0.5ex]
\colhead{quantity}	&	\colhead{($^\circ$)}	&	\colhead{}	&	\colhead{}	&	\colhead{}	&	\colhead{(\%)}	&	\colhead{}	&	\colhead{}	&	\colhead{(\%)}	&	\colhead{}	&	\colhead{}	&	\colhead{(\%)}	&	\colhead{}	&	\colhead{}	&	\colhead{(\%)}	&	\colhead{}	&	\colhead{}	&	\colhead{}
}
\startdata
$T_{90}$	&	20	&	100	&	0.17	&	100	&	51	&	0.19	&	50	&	47	&	0.24	&	12	&	31	&	0.29	&	1	&	40	&	0.41	&	0	&	0.30	\\[-1.0ex]
$T_{90}$	&	30	&	100	&	0.11	&	68	&	80	&	0.13	&	33	&	73	&	0.15	&	6	&	56	&	0.18	&	0	&	100	&	0.25	&	0	&	0.18	\\[-1.0ex]
$T_{90}$	&	40	&	100	&	0.08	&	59	&	69	&	0.09	&	26	&	68	&	0.11	&	0	&	100	&	0.14	&	0	&	100	&	0.17	&	0	&	0.11	\\[-1.0ex]
$T_{90}$	&	50	&	100	&	0.07	&	83	&	50	&	0.07	&	28	&	58	&	0.09	&	0	&	100	&	0.11	&	0	&	100	&	0.14	&	0	&	0.09	\\[-1.0ex]
$T_{90}$	&	60	&	100	&	0.05	&	57	&	65	&	0.06	&	15	&	65	&	0.07	&	1	&	59	&	0.08	&	0	&	100	&	0.10	&	0	&	0.07	\\[-1.0ex]
$S$	&	20	&	100	&	0.17	&	114	&	28	&	0.19	&	72	&	14	&	0.23	&	34	&	{\bf 1}	&	0.28	&	9	&	{\bf 2}	&	0.38	&	1	&	0.40	\\[-1.0ex]
$S$	&	20	&	1000	&	0.17	&	113	&	30.2	&	0.19	&	72	&	13.1	&	0.23	&	30	&	{\bf 2.2}	&	0.28	&	9	&	{\bf 1.1}	&	0.42	&	0	&	0.40	\\[-1.0ex]
$S$	&	30	&	100	&	0.11	&	89	&	58	&	0.13	&	60	&	29	&	0.15	&	23	&	13	&	0.19	&	3	&	9	&	0.24	&	0	&	0.20	\\[-1.0ex]
$S$	&	40	&	100	&	0.08	&	95	&	49	&	0.09	&	50	&	40	&	0.11	&	4	&	57	&	0.13	&	0	&	100	&	0.15	&	0	&	0.12	\\[-1.0ex]
$S$	&	50	&	100	&	0.07	&	66	&	71	&	0.07	&	24	&	73	&	0.09	&	4	&	49	&	0.10	&	0	&	100	&	0.12	&	0	&	0.09	\\[-1.0ex]
$S$	&	60	&	100	&	0.05	&	52	&	64	&	0.06	&	18	&	62	&	0.07	&	1	&	53	&	0.09	&	0	&	100	&	0.11	&	0	&	0.07	\\[-1.0ex]
$S_\mathrm{B}$	&	20	&	100	&	0.17	&	123	&	22	&	0.19	&	72	&	15	&	0.23	&	30	&	{\bf $<$1}	&	0.28	&	5	&	6	&	0.36	&	1	&	0.36	\\[-1.0ex]
$S_\mathrm{B}$	&	20	&	1000	&	0.17	&	124	&	17.3	&	0.19	&	74	&	11.5	&	0.23	&	31	&	{\bf 1.4}	&	0.29	&	3	&	12.8	&	0.47	&	0	&	0.36	\\[-1.0ex]
$S_\mathrm{B}$	&	30	&	100	&	0.11	&	119	&	32	&	0.13	&	58	&	39	&	0.15	&	21	&	13	&	0.18	&	2	&	19	&	0.22	&	0	&	0.19	\\[-1.0ex]
$S_\mathrm{B}$	&	40	&	100	&	0.08	&	79	&	56	&	0.09	&	46	&	44	&	0.11	&	1	&	80	&	0.14	&	0	&	100	&	0.19	&	0	&	0.12	\\[-1.0ex]
$S_\mathrm{B}$	&	50	&	100	&	0.06	&	93	&	49	&	0.07	&	32	&	60	&	0.08	&	0	&	100	&	0.10	&	0	&	100	&	0.13	&	0	&	0.08	\\[-1.0ex]
$S_\mathrm{B}$	&	60	&	100	&	0.05	&	97	&	37	&	0.06	&	36	&	48	&	0.07	&	2	&	55	&	0.08	&	0	&	100	&	0.10	&	0	&	0.07	\\[-1.0ex]
$F_{64}$	&	20	&	100	&	0.17	&	145	&	8	&	0.19	&	87	&	{\bf 4}	&	0.23	&	34	&	{\bf 2}	&	0.28	&	15	&	{\bf $<$1}	&	0.38	&	1	&	0.39	\\[-1.0ex]
$F_{64}$	&	20	&	1000	&	0.17	&	144	&	5.6	&	0.19	&	87	&	{\bf 3.4}	&	0.23	&	30	&	{\bf 1.5}	&	0.28	&	14	&	{\bf 0.4}	&	0.42	&	0	&	0.39	\\[-1.0ex]
$F_{64}$	&	30	&	100	&	0.11	&	124	&	27	&	0.13	&	74	&	19	&	0.15	&	33	&	{\bf 3}	&	0.18	&	13	&	{\bf 1}	&	0.24	&	0	&	0.22	\\[-1.0ex]
$F_{64}$	&	30	&	1000	&	0.11	&	117	&	31.7	&	0.13	&	69	&	24.2	&	0.15	&	30	&	6.4	&	0.19	&	12	&	{\bf 1.2}	&	0.27	&	0	&	0.22	\\[-1.0ex]
$F_{64}$	&	40	&	100	&	0.08	&	132	&	25	&	0.09	&	89	&	14	&	0.11	&	32	&	8	&	0.14	&	4	&	9	&	0.17	&	0	&	0.14	\\[-1.0ex]
$F_{64}$	&	50	&	100	&	0.07	&	134	&	27	&	0.07	&	61	&	34	&	0.09	&	14	&	23	&	0.11	&	1	&	20	&	0.14	&	0	&	0.11	\\[-1.0ex]
$F_{64}$	&	60	&	100	&	0.05	&	143	&	21	&	0.06	&	75	&	22	&	0.07	&	17	&	20	&	0.08	&	0	&	100	&	0.10	&	0	&	0.08	\\[-1.0ex]
$F_{256}$	&	20	&	100	&	0.17	&	109	&	37	&	0.19	&	60	&	30	&	0.23	&	17	&	18	&	0.28	&	6	&	{\bf 3}	&	0.36	&	1	&	0.40	\\[-1.0ex]
$F_{256}$	&	20	&	1000	&	0.17	&	107	&	38.9	&	0.19	&	58	&	31.7	&	0.23	&	17	&	18.3	&	0.29	&	6	&	{\bf 3.5}	&	0.41	&	0	&	0.40	\\[-1.0ex]
$F_{256}$	&	30	&	100	&	0.11	&	108	&	36	&	0.13	&	69	&	23	&	0.15	&	26	&	12	&	0.18	&	8	&	{\bf 2}	&	0.23	&	0	&	0.23	\\[-1.0ex]
$F_{256}$	&	30	&	1000	&	0.11	&	103	&	43.8	&	0.13	&	65	&	26.8	&	0.15	&	25	&	9.6	&	0.18	&	8	&	{\bf 3.6}	&	0.25	&	0	&	0.23	\\[-1.0ex]
$F_{256}$	&	40	&	100	&	0.08	&	140	&	23	&	0.09	&	87	&	17	&	0.11	&	31	&	10	&	0.14	&	2	&	11	&	0.20	&	0	&	0.15	\\[-1.0ex]
$F_{256}$	&	50	&	100	&	0.07	&	143	&	23	&	0.07	&	76	&	25	&	0.09	&	27	&	12	&	0.11	&	6	&	7	&	0.14	&	0	&	0.12	\\[-1.0ex]
$F_{256}$	&	60	&	100	&	0.05	&	166	&	25	&	0.06	&	93	&	22	&	0.07	&	35	&	11	&	0.08	&	2	&	11	&	0.10	&	0	&	0.09	\\[-1.0ex]
$F_{1024}$	&	20	&	100	&	0.17	&	78	&	76	&	0.19	&	43	&	63	&	0.23	&	13	&	28	&	0.28	&	6	&	{\bf 2}	&	0.36	&	1	&	0.37	\\[-1.0ex]
$F_{1024}$	&	20	&	1000	&	0.17	&	74	&	84.8	&	0.19	&	40	&	69.9	&	0.23	&	13	&	29.6	&	0.28	&	6	&	{\bf 2.9}	&	0.44	&	0	&	0.37	\\[-1.0ex]
$F_{1024}$	&	30	&	100	&	0.11	&	68	&	77	&	0.13	&	33	&	70	&	0.15	&	11	&	31	&	0.18	&	1	&	26	&	0.23	&	0	&	0.20	\\[-1.0ex]
$F_{1024}$	&	40	&	100	&	0.08	&	51	&	81	&	0.09	&	27	&	66	&	0.11	&	2	&	65	&	0.14	&	0	&	100	&	0.18	&	0	&	0.12	\\[-1.0ex]
$F_{1024}$	&	50	&	100	&	0.07	&	83	&	48	&	0.07	&	35	&	48	&	0.09	&	7	&	31	&	0.10	&	0	&	100	&	0.13	&	0	&	0.10	\\[-1.0ex]
$F_{1024}$	&	60	&	100	&	0.05	&	78	&	49	&	0.06	&	41	&	43	&	0.07	&	6	&	43	&	0.08	&	0	&	100	&	0.09	&	0	&	0.07	\\[-1.0ex]
$F_{64\mathrm{,B}}$	&	20	&	100	&	0.17	&	131	&	13	&	0.19	&	78	&	{\bf 4}	&	0.23	&	28	&	{\bf 2}	&	0.28	&	7	&	{\bf 3}	&	0.41	&	0	&	0.32	\\[-1.0ex]
$F_{64\mathrm{,B}}$	&	20	&	1000	&	0.17	&	129	&	15.7	&	0.19	&	78	&	8.9	&	0.23	&	28	&	{\bf 2.7}	&	0.28	&	6	&	{\bf 3.4}	&	0.42	&	0	&	0.32	\\[-1.0ex]
$F_{64\mathrm{,B}}$	&	30	&	100	&	0.11	&	123	&	26	&	0.13	&	74	&	20	&	0.15	&	25	&	11	&	0.19	&	3	&	9	&	0.24	&	0	&	0.20	\\[-1.0ex]
$F_{64\mathrm{,B}}$	&	30	&	1000	&	0.11	&	125	&	26.2	&	0.13	&	75	&	18.2	&	0.15	&	25	&	9.1	&	0.18	&	5	&	6.6	&	0.26	&	0	&	0.20	\\[-1.0ex]
$F_{64\mathrm{,B}}$	&	40	&	100	&	0.08	&	131	&	29	&	0.09	&	82	&	18	&	0.11	&	21	&	16	&	0.13	&	3	&	14	&	0.17	&	0	&	0.15	\\[-1.0ex]
$F_{64\mathrm{,B}}$	&	40	&	1000	&	0.08	&	130	&	26.2	&	0.09	&	78	&	19.8	&	0.11	&	21	&	15.8	&	0.13	&	3	&	10.0	&	0.21	&	0	&	0.15	\\[-1.0ex]
$F_{64\mathrm{,B}}$	&	50	&	100	&	0.07	&	107	&	40	&	0.08	&	48	&	46	&	0.09	&	0	&	100	&	0.11	&	0	&	100	&	0.17	&	0	&	0.09	\\[-1.0ex]
$F_{64\mathrm{,B}}$	&	60	&	100	&	0.05	&	109	&	38	&	0.06	&	41	&	47	&	0.07	&	5	&	41	&	0.08	&	0	&	100	&	0.10	&	0	&	0.07	\\[-1.0ex]
$F_{256\mathrm{,B}}$	&	20	&	100	&	0.17	&	103	&	48	&	0.19	&	60	&	34	&	0.23	&	26	&	{\bf 5}	&	0.28	&	5	&	{\bf 3}	&	0.37	&	0	&	0.32	\\[-1.0ex]
$F_{256\mathrm{,B}}$	&	20	&	1000	&	0.17	&	99	&	49.7	&	0.19	&	60	&	27.2	&	0.23	&	23	&	6.4	&	0.28	&	5	&	5.8	&	0.41	&	0	&	0.32	\\[-1.0ex]
$F_{256\mathrm{,B}}$	&	30	&	100	&	0.11	&	99	&	53	&	0.13	&	66	&	27	&	0.15	&	13	&	29	&	0.19	&	0	&	100	&	0.23	&	0	&	0.19	\\[-1.0ex]
$F_{256\mathrm{,B}}$	&	40	&	100	&	0.08	&	74	&	62	&	0.09	&	45	&	44	&	0.11	&	7	&	39	&	0.14	&	0	&	100	&	0.19	&	0	&	0.14	\\[-1.0ex]
$F_{256\mathrm{,B}}$	&	50	&	100	&	0.07	&	71	&	63	&	0.07	&	24	&	63	&	0.09	&	2	&	55	&	0.11	&	0	&	100	&	0.13	&	0	&	0.09	\\[-1.0ex]
$F_{256\mathrm{,B}}$	&	60	&	100	&	0.05	&	82	&	46	&	0.06	&	42	&	39	&	0.07	&	4	&	34	&	0.09	&	0	&	100	&	0.11	&	0	&	0.07	\\[-1.0ex]
$F_{1024\mathrm{,B}}$	&	20	&	100	&	0.17	&	90	&	66	&	0.19	&	58	&	34	&	0.23	&	8	&	58	&	0.29	&	1	&	39	&	0.37	&	0	&	0.31	\\[-1.0ex]
$F_{1024\mathrm{,B}}$	&	30	&	100	&	0.11	&	98	&	50	&	0.13	&	53	&	37	&	0.15	&	8	&	43	&	0.18	&	0	&	100	&	0.25	&	0	&	0.17	\\[-1.0ex]
$F_{1024\mathrm{,B}}$	&	40	&	100	&	0.09	&	84	&	48	&	0.09	&	40	&	47	&	0.12	&	1	&	73	&	0.14	&	0	&	100	&	0.17	&	0	&	0.12	\\[-1.0ex]
$F_{1024\mathrm{,B}}$	&	50	&	100	&	0.07	&	111	&	35	&	0.07	&	52	&	35	&	0.09	&	6	&	33	&	0.11	&	0	&	100	&	0.14	&	0	&	0.09	\\[-1.0ex]
$F_{1024\mathrm{,B}}$	&	60	&	100	&	0.05	&	147	&	25	&	0.06	&	63	&	31	&	0.07	&	8	&	36	&	0.08	&	0	&	100	&	0.10	&	0	&	0.07	\\
\enddata
\tablenotetext{\ast}{The radii of the patches.\\[-3.5ex]}
\tablenotetext{\star}{The number of random shufflings of the data.\\[-3.5ex]}
\tablenotetext{\dagger}{$D^\mathrm{s}_{10}$, $D^\mathrm{s}_{5}$, $D^\mathrm{s}_{1}$,
and $D^\mathrm{s}_{0.1}$ delimit the highest 10\,\%, 5\,\%, 1\,\%, and 0.1\,\% of all
$D^\mathrm{s}$ values from all patches in all randomly shuffled data, respectively.
$D^\mathrm{s}_0$ is the maximum of all $D^\mathrm{s}$ values in all patches of all shuffled data.\\[-3.5ex]}
\tablenotetext{\ddagger}{$N^\mathrm{m}_\mathrm{i}$ is the number of patches in the
measured data for which $D^\mathrm{m}>D^\mathrm{s}_\mathrm{i}$, where i=10, 5, 1, or 0.1.
$N^\mathrm{m}_0$ is the number of patches in the measured data with $D^\mathrm{m}>D^\mathrm{s}_\mathrm{0}$.\\[-3.5ex]}
\tablenotetext{\#}{$P^\mathrm{N}_\mathrm{i}$ is the probability of finding at least
$N^\mathrm{m}_\mathrm{i}$ number of patches with $D^\mathrm{s}>D^\mathrm{s}_\mathrm{i}$
in the randomly shuffled data, where i=10, 5, 1, or 0.1. The cases with
$P^\mathrm{N}_\mathrm{i}\leq 5$\,\% are emphasized in boldface.\\[-3.5ex]}
\tablenotetext{\S}{$D^\mathrm{m}_{0}$ is the maximum value of the statistic in the measured data.\\[-3.5ex]}
\end{deluxetable*}

\begin{deluxetable*}{cccccccccccccccccc}
\tablecaption{\label{tab:results_kuiper}
Results using the Kuiper statistic $V$.}
\tablewidth{0pt}
\tabletypesize{\scriptsize}
\tablehead{
\colhead{Tested}	&	\colhead{$r$\tablenotemark{$\ast$}}	&	\colhead{$n$\tablenotemark{$\star$}}	&	\colhead{$V^\mathrm{s}_{10}$\tablenotemark{$\dagger$}}	&	\colhead{$N^\mathrm{m}_{10}$\tablenotemark{$\ddagger$}}	&	\colhead{$P^\mathrm{N}_{10}$\tablenotemark{$\#$}}	&	\colhead{$V^\mathrm{s}_{5}$\tablenotemark{$\dagger$}}	&	\colhead{$N^\mathrm{m}_{5}$\tablenotemark{$\ddagger$}}	&	\colhead{$P^\mathrm{N}_{5}$\tablenotemark{$\#$}}	&	\colhead{$V^\mathrm{s}_{1}$\tablenotemark{$\dagger$}}	&	\colhead{$N^\mathrm{m}_{1}$\tablenotemark{$\ddagger$}}	&	\colhead{$P^\mathrm{N}_{1}$\tablenotemark{$\#$}}	&	\colhead{$V^\mathrm{s}_{0.1}$\tablenotemark{$\dagger$}}	&	\colhead{$N^\mathrm{m}_{0.1}$\tablenotemark{$\ddagger$}}	&	\colhead{$P^\mathrm{N}_{0.1}$\tablenotemark{$\#$}}	&
\colhead{$V^\mathrm{s}_0$\tablenotemark{$\dagger$}}	&	\colhead{$N^\mathrm{m}_0$\tablenotemark{$\ddagger$}}	&	\colhead{$V^\mathrm{m}_0$\tablenotemark{\S}}
\\[-0.5ex]
\colhead{quantity}	&	\colhead{($^\circ$)}	&	\colhead{}	&	\colhead{}	&	\colhead{}	&	\colhead{(\%)}	&	\colhead{}	&	\colhead{}	&	\colhead{(\%)}	&	\colhead{}	&	\colhead{}	&	\colhead{(\%)}	&	\colhead{}	&	\colhead{}	&	\colhead{(\%)}	&	\colhead{}	&	\colhead{}	&	\colhead{}
}
\startdata
$T_{90}$	&	20	&	100	&	0.23	&	120	&	23	&	0.25	&	57	&	32	&	0.29	&	14	&	25	&	0.34	&	0	&	100	&	0.41	&	0	&	0.33	\\[-1.0ex]
$T_{90}$	&	30	&	100	&	0.15	&	138	&	14	&	0.16	&	82	&	9	&	0.19	&	12	&	32	&	0.22	&	0	&	100	&	0.27	&	0	&	0.21	\\[-1.0ex]
$T_{90}$	&	40	&	100	&	0.11	&	112	&	36	&	0.12	&	41	&	57	&	0.14	&	7	&	44	&	0.16	&	0	&	100	&	0.19	&	0	&	0.15	\\[-1.0ex]
$T_{90}$	&	50	&	100	&	0.09	&	109	&	36	&	0.09	&	37	&	55	&	0.11	&	3	&	58	&	0.12	&	0	&	100	&	0.14	&	0	&	0.11	\\[-1.0ex]
$T_{90}$	&	60	&	100	&	0.07	&	115	&	36	&	0.08	&	46	&	43	&	0.09	&	5	&	46	&	0.10	&	0	&	100	&	0.12	&	0	&	0.09	\\[-1.0ex]
$S$	&	20	&	100	&	0.23	&	96	&	59	&	0.25	&	63	&	22	&	0.28	&	23	&	9	&	0.34	&	6	&	{\bf 5}	&	0.40	&	1	&	0.41	\\[-1.0ex]
$S$	&	20	&	1000	&	0.23	&	95	&	55.5	&	0.25	&	62	&	23.7	&	0.29	&	23	&	5.6	&	0.33	&	6	&	{\bf 3.3}	&	0.51	&	0	&	0.41	\\[-1.0ex]
$S$	&	30	&	100	&	0.15	&	87	&	66	&	0.16	&	50	&	49	&	0.19	&	15	&	21	&	0.22	&	0	&	100	&	0.27	&	0	&	0.22	\\[-1.0ex]
$S$	&	40	&	100	&	0.11	&	61	&	71	&	0.12	&	29	&	68	&	0.14	&	6	&	48	&	0.16	&	0	&	100	&	0.21	&	0	&	0.15	\\[-1.0ex]
$S$	&	50	&	100	&	0.09	&	24	&	91	&	0.09	&	5	&	93	&	0.11	&	0	&	100	&	0.12	&	0	&	100	&	0.14	&	0	&	0.10	\\[-1.0ex]
$S$	&	60	&	100	&	0.07	&	15	&	95	&	0.08	&	2	&	98	&	0.09	&	0	&	100	&	0.10	&	0	&	100	&	0.12	&	0	&	0.08	\\[-1.0ex]
$S_\mathrm{B}$	&	20	&	100	&	0.23	&	110	&	33	&	0.25	&	68	&	15	&	0.29	&	23	&	6	&	0.34	&	3	&	14	&	0.42	&	0	&	0.37	\\[-1.0ex]
$S_\mathrm{B}$	&	20	&	1000	&	0.23	&	108	&	35.6	&	0.25	&	66	&	18.0	&	0.29	&	23	&	5.9	&	0.34	&	3	&	13.2	&	0.49	&	0	&	0.37	\\[-1.0ex]
$S_\mathrm{B}$	&	30	&	100	&	0.15	&	101	&	52	&	0.16	&	54	&	43	&	0.19	&	20	&	12	&	0.22	&	1	&	40	&	0.28	&	0	&	0.22	\\[-1.0ex]
$S_\mathrm{B}$	&	40	&	100	&	0.11	&	57	&	70	&	0.12	&	29	&	62	&	0.14	&	3	&	58	&	0.16	&	0	&	100	&	0.19	&	0	&	0.15	\\[-1.0ex]
$S_\mathrm{B}$	&	50	&	100	&	0.08	&	42	&	87	&	0.09	&	10	&	95	&	0.10	&	0	&	100	&	0.12	&	0	&	100	&	0.15	&	0	&	0.10	\\[-1.0ex]
$S_\mathrm{B}$	&	60	&	100	&	0.07	&	22	&	91	&	0.07	&	5	&	92	&	0.08	&	0	&	100	&	0.10	&	0	&	100	&	0.11	&	0	&	0.08	\\[-1.0ex]
$F_{64}$	&	20	&	100	&	0.23	&	141	&	{\bf 5}	&	0.25	&	77	&	9	&	0.28	&	27	&	{\bf 2}	&	0.33	&	7	&	{\bf 2}	&	0.42	&	0	&	0.40	\\[-1.0ex]
$F_{64}$	&	20	&	1000	&	0.23	&	140	&	6.3	&	0.25	&	77	&	6.6	&	0.29	&	27	&	{\bf 2.4}	&	0.33	&	7	&	{\bf 1.9}	&	0.45	&	0	&	0.40	\\[-1.0ex]
$F_{64}$	&	30	&	100	&	0.15	&	112	&	38	&	0.16	&	66	&	28	&	0.19	&	24	&	11	&	0.21	&	5	&	7	&	0.26	&	0	&	0.22	\\[-1.0ex]
$F_{64}$	&	30	&	1000	&	0.15	&	112	&	35.5	&	0.16	&	68	&	24.6	&	0.19	&	24	&	9.9	&	0.22	&	4	&	8.2	&	0.28	&	0	&	0.22	\\[-1.0ex]
$F_{64}$	&	40	&	100	&	0.11	&	116	&	28	&	0.12	&	64	&	25	&	0.14	&	22	&	17	&	0.16	&	3	&	14	&	0.18	&	0	&	0.16	\\[-1.0ex]
$F_{64}$	&	50	&	100	&	0.09	&	121	&	33	&	0.09	&	51	&	38	&	0.11	&	4	&	48	&	0.12	&	0	&	100	&	0.16	&	0	&	0.11	\\[-1.0ex]
$F_{64}$	&	60	&	100	&	0.07	&	119	&	28	&	0.08	&	50	&	38	&	0.09	&	1	&	66	&	0.10	&	0	&	100	&	0.13	&	0	&	0.09	\\[-1.0ex]
$F_{256}$	&	20	&	100	&	0.23	&	110	&	29	&	0.25	&	53	&	40	&	0.29	&	14	&	28	&	0.33	&	4	&	10	&	0.41	&	1	&	0.43	\\[-1.0ex]
$F_{256}$	&	20	&	1000	&	0.23	&	110	&	34.7	&	0.25	&	52	&	42.6	&	0.29	&	14	&	25.4	&	0.34	&	4	&	8.5	&	0.48	&	0	&	0.43	\\[-1.0ex]
$F_{256}$	&	30	&	100	&	0.15	&	107	&	36	&	0.16	&	62	&	29	&	0.19	&	18	&	19	&	0.22	&	1	&	29	&	0.26	&	0	&	0.25	\\[-1.0ex]
$F_{256}$	&	30	&	1000	&	0.15	&	106	&	40.5	&	0.16	&	61	&	30.1	&	0.19	&	18	&	17.5	&	0.22	&	1	&	32.4	&	0.26	&	0	&	0.25	\\[-1.0ex]
$F_{256}$	&	40	&	100	&	0.11	&	135	&	23	&	0.12	&	75	&	24	&	0.14	&	20	&	15	&	0.16	&	2	&	10	&	0.21	&	0	&	0.17	\\[-1.0ex]
$F_{256}$	&	50	&	100	&	0.09	&	128	&	28	&	0.09	&	67	&	30	&	0.11	&	15	&	21	&	0.13	&	1	&	18	&	0.15	&	0	&	0.13	\\[-1.0ex]
$F_{256}$	&	60	&	100	&	0.07	&	132	&	27	&	0.07	&	70	&	25	&	0.08	&	19	&	18	&	0.10	&	0	&	100	&	0.12	&	0	&	0.10	\\[-1.0ex]
$F_{1024}$	&	20	&	100	&	0.23	&	76	&	78	&	0.25	&	48	&	56	&	0.28	&	13	&	30	&	0.33	&	6	&	{\bf 4}	&	0.41	&	1	&	0.42	\\[-1.0ex]
$F_{1024}$	&	20	&	1000	&	0.23	&	76	&	82.0	&	0.25	&	48	&	52.6	&	0.29	&	13	&	30.5	&	0.34	&	5	&	{\bf 4.7}	&	0.50	&	0	&	0.42	\\[-1.0ex]
$F_{1024}$	&	30	&	100	&	0.15	&	53	&	88	&	0.16	&	36	&	65	&	0.19	&	8	&	44	&	0.22	&	1	&	30	&	0.25	&	0	&	0.24	\\[-1.0ex]
$F_{1024}$	&	40	&	100	&	0.11	&	35	&	92	&	0.12	&	12	&	92	&	0.14	&	0	&	100	&	0.16	&	0	&	100	&	0.19	&	0	&	0.14	\\[-1.0ex]
$F_{1024}$	&	50	&	100	&	0.09	&	47	&	86	&	0.09	&	17	&	85	&	0.11	&	1	&	83	&	0.12	&	0	&	100	&	0.14	&	0	&	0.11	\\[-1.0ex]
$F_{1024}$	&	60	&	100	&	0.07	&	27	&	85	&	0.07	&	5	&	92	&	0.08	&	0	&	100	&	0.10	&	0	&	100	&	0.11	&	0	&	0.08	\\[-1.0ex]
$F_{64\mathrm{,B}}$	&	20	&	100	&	0.23	&	141	&	10	&	0.25	&	72	&	14	&	0.29	&	23	&	8	&	0.34	&	3	&	12	&	0.50	&	0	&	0.34	\\[-1.0ex]
$F_{64\mathrm{,B}}$	&	20	&	1000	&	0.23	&	137	&	9.2	&	0.25	&	67	&	17.2	&	0.29	&	23	&	5.8	&	0.33	&	3	&	13.7	&	0.48	&	0	&	0.34	\\[-1.0ex]
$F_{64\mathrm{,B}}$	&	30	&	100	&	0.15	&	137	&	16	&	0.16	&	84	&	12	&	0.19	&	24	&	10	&	0.22	&	1	&	34	&	0.31	&	0	&	0.22	\\[-1.0ex]
$F_{64\mathrm{,B}}$	&	30	&	1000	&	0.15	&	137	&	17.7	&	0.16	&	81	&	13.7	&	0.19	&	24	&	9.7	&	0.22	&	1	&	31.5	&	0.28	&	0	&	0.22	\\[-1.0ex]
$F_{64\mathrm{,B}}$	&	40	&	100	&	0.11	&	119	&	37	&	0.12	&	78	&	22	&	0.14	&	35	&	9	&	0.16	&	9	&	{\bf 4}	&	0.19	&	0	&	0.18	\\[-1.0ex]
$F_{64\mathrm{,B}}$	&	40	&	1000	&	0.11	&	121	&	31.1	&	0.12	&	78	&	17.8	&	0.14	&	35	&	{\bf 4.7}	&	0.16	&	9	&	{\bf 2.5}	&	0.23	&	0	&	0.18	\\[-1.0ex]
$F_{64\mathrm{,B}}$	&	50	&	100	&	0.09	&	91	&	46	&	0.10	&	54	&	34	&	0.11	&	11	&	28	&	0.13	&	0	&	100	&	0.17	&	0	&	0.13	\\[-1.0ex]
$F_{64\mathrm{,B}}$	&	60	&	100	&	0.07	&	52	&	77	&	0.08	&	26	&	71	&	0.09	&	9	&	38	&	0.10	&	0	&	100	&	0.12	&	0	&	0.10	\\[-1.0ex]
$F_{256\mathrm{,B}}$	&	20	&	100	&	0.23	&	105	&	41	&	0.25	&	68	&	19	&	0.28	&	22	&	10	&	0.33	&	1	&	43	&	0.44	&	0	&	0.34	\\[-1.0ex]
$F_{256\mathrm{,B}}$	&	20	&	1000	&	0.23	&	103	&	43.0	&	0.25	&	64	&	20.9	&	0.29	&	21	&	7.3	&	0.33	&	1	&	43.9	&	0.46	&	0	&	0.34	\\[-1.0ex]
$F_{256\mathrm{,B}}$	&	30	&	100	&	0.15	&	74	&	71	&	0.16	&	40	&	62	&	0.19	&	13	&	27	&	0.21	&	0	&	100	&	0.27	&	0	&	0.21	\\[-1.0ex]
$F_{256\mathrm{,B}}$	&	40	&	100	&	0.11	&	66	&	72	&	0.12	&	43	&	50	&	0.14	&	9	&	36	&	0.16	&	0	&	100	&	0.24	&	0	&	0.15	\\[-1.0ex]
$F_{256\mathrm{,B}}$	&	50	&	100	&	0.09	&	54	&	76	&	0.09	&	18	&	79	&	0.11	&	2	&	65	&	0.12	&	0	&	100	&	0.16	&	0	&	0.11	\\[-1.0ex]
$F_{256\mathrm{,B}}$	&	60	&	100	&	0.07	&	58	&	63	&	0.07	&	30	&	59	&	0.09	&	7	&	36	&	0.10	&	0	&	100	&	0.12	&	0	&	0.09	\\[-1.0ex]
$F_{1024\mathrm{,B}}$	&	20	&	100	&	0.23	&	98	&	49	&	0.25	&	52	&	43	&	0.29	&	15	&	19	&	0.33	&	0	&	100	&	0.46	&	0	&	0.32	\\[-1.0ex]
$F_{1024\mathrm{,B}}$	&	30	&	100	&	0.15	&	102	&	40	&	0.16	&	59	&	33	&	0.19	&	8	&	49	&	0.22	&	0	&	100	&	0.28	&	0	&	0.22	\\[-1.0ex]
$F_{1024\mathrm{,B}}$	&	40	&	100	&	0.11	&	111	&	37	&	0.12	&	47	&	44	&	0.14	&	2	&	75	&	0.16	&	0	&	100	&	0.20	&	0	&	0.15	\\[-1.0ex]
$F_{1024\mathrm{,B}}$	&	50	&	100	&	0.09	&	138	&	26	&	0.09	&	58	&	33	&	0.11	&	7	&	42	&	0.12	&	0	&	100	&	0.15	&	0	&	0.11	\\[-1.0ex]
$F_{1024\mathrm{,B}}$	&	60	&	100	&	0.07	&	122	&	33	&	0.07	&	58	&	32	&	0.09	&	1	&	67	&	0.10	&	0	&	100	&	0.11	&	0	&	0.09	\\
\enddata
\tablenotetext{\ast}{The radii of the patches.\\[-3.5ex]}
\tablenotetext{\star}{The number of random shufflings of the data.\\[-3.5ex]}
\tablenotetext{\dagger}{$V^\mathrm{s}_{10}$, $V^\mathrm{s}_{5}$, $V^\mathrm{s}_{1}$,
and $V^\mathrm{s}_{0.1}$ delimit the highest 10\,\%, 5\,\%, 1\,\%, and 0.1\,\% of all
$V^\mathrm{s}$ values from all patches in all randomly shuffled data, respectively.
$V^\mathrm{s}_0$ is the maximum of all $V^\mathrm{s}$ values in all patches of all shuffled data.\\[-3.5ex]}
\tablenotetext{\ddagger}{$N^\mathrm{m}_\mathrm{i}$ is the number of patches in the
measured data for which $V^\mathrm{m}>V^\mathrm{s}_\mathrm{i}$, where i=10, 5, 1, or 0.1.
$N^\mathrm{m}_0$ is the number of patches in the measured data with $V^\mathrm{m}>V^\mathrm{s}_\mathrm{0}$.\\[-3.5ex]}
\tablenotetext{\#}{$P^\mathrm{N}_\mathrm{i}$ is the probability of finding at least
$N^\mathrm{m}_\mathrm{i}$ number of patches with $V^\mathrm{s}>V^\mathrm{s}_\mathrm{i}$
in the randomly shuffled data, where i=10, 5, 1, or 0.1. The cases with
$P^\mathrm{N}_\mathrm{i}\leq 5$\,\% are emphasized in boldface.\\[-3.5ex]}
\tablenotetext{\S}{$V^\mathrm{m}_{0}$ is the maximum value of the statistic in the measured data.}
\end{deluxetable*}

\newpage

\begin{deluxetable*}{cccccccccccccccccc}
\tablecaption{\label{tab:results_ad}
Results using the Anderson--Darling statistic $AD$.}
\tablewidth{0pt}
\tabletypesize{\scriptsize}
\tablehead{
\colhead{Tested}	&	\colhead{$r$\tablenotemark{$\ast$}}	&	\colhead{$n$\tablenotemark{$\star$}}	&	\colhead{$AD^\mathrm{s}_{10}$\tablenotemark{$\dagger$}}	&	\colhead{$N^\mathrm{m}_{10}$\tablenotemark{$\ddagger$}}	&	\colhead{$P^\mathrm{N}_{10}$\tablenotemark{$\#$}}	&	\colhead{$AD^\mathrm{s}_{5}$\tablenotemark{$\dagger$}}	&	\colhead{$N^\mathrm{m}_{5}$\tablenotemark{$\ddagger$}}	&	\colhead{$P^\mathrm{N}_{5}$\tablenotemark{$\#$}}	&	\colhead{$AD^\mathrm{s}_{1}$\tablenotemark{$\dagger$}}	&	\colhead{$N^\mathrm{m}_{1}$\tablenotemark{$\ddagger$}}	&	\colhead{$P^\mathrm{N}_{1}$\tablenotemark{$\#$}}	&	\colhead{$AD^\mathrm{s}_{0.1}$\tablenotemark{$\dagger$}}	&	\colhead{$N^\mathrm{m}_{0.1}$\tablenotemark{$\ddagger$}}	&	\colhead{$P^\mathrm{N}_{0.1}$\tablenotemark{$\#$}}	&	\colhead{$AD^\mathrm{s}_0$\tablenotemark{$\dagger$}}	&	\colhead{$N^\mathrm{m}_0$\tablenotemark{$\ddagger$}}	&	\colhead{$AD^\mathrm{m}_0$\tablenotemark{\S}}
\\[-0.5ex]
\colhead{quantity}	&	\colhead{($^\circ$)}	&	\colhead{}	&	\colhead{}	&	\colhead{}	&	\colhead{(\%)}	&	\colhead{}	&	\colhead{}	&	\colhead{(\%)}	&	\colhead{}	&	\colhead{}	&	\colhead{(\%)}	&	\colhead{}	&	\colhead{}	&	\colhead{(\%)}	&	\colhead{}	&	\colhead{}	&	\colhead{}
}
\startdata
$T_{90}$	&	20	&	100	&	1.80	&	99	&	50	&	2.32	&	32	&	79	&	3.61	&	5	&	71	&	5.82	&	0	&	100	&	14.00	&	0	&	4.41	\\[-1.0ex]
$T_{90}$	&	30	&	100	&	1.65	&	70	&	75	&	2.12	&	31	&	70	&	3.28	&	1	&	92	&	5.35	&	0	&	100	&	9.96	&	0	&	3.31	\\[-1.0ex]
$T_{90}$	&	40	&	100	&	1.57	&	57	&	73	&	2.04	&	21	&	70	&	3.16	&	1	&	68	&	4.92	&	0	&	100	&	7.94	&	0	&	3.18	\\[-1.0ex]
$T_{90}$	&	50	&	100	&	1.33	&	68	&	62	&	1.72	&	23	&	60	&	2.72	&	1	&	56	&	4.31	&	0	&	100	&	6.89	&	0	&	2.97	\\[-1.0ex]
$T_{90}$	&	60	&	100	&	1.20	&	49	&	67	&	1.54	&	19	&	65	&	2.27	&	0	&	100	&	3.32	&	0	&	100	&	4.89	&	0	&	2.19	\\[-1.0ex]
$S$	&	20	&	100	&	1.81	&	117	&	29	&	2.33	&	74	&	12	&	3.61	&	24	&	{\bf 3}	&	5.45	&	1	&	36	&	10.33	&	0	&	7.74	\\[-1.0ex]
$S$	&	20	&	1000	&	1.81	&	117	&	28.5	&	2.33	&	74	&	12.7	&	3.62	&	24	&	7.0	&	5.56	&	1	&	35.7	&	14.90	&	0	&	7.74	\\[-1.0ex]
$S$	&	30	&	100	&	1.71	&	90	&	51	&	2.22	&	56	&	35	&	3.45	&	15	&	24	&	5.34	&	0	&	100	&	8.93	&	0	&	4.70	\\[-1.0ex]
$S$	&	40	&	100	&	1.50	&	80	&	57	&	1.94	&	25	&	71	&	3.00	&	2	&	70	&	4.45	&	0	&	100	&	6.95	&	0	&	3.62	\\[-1.0ex]
$S$	&	50	&	100	&	1.37	&	70	&	64	&	1.75	&	26	&	69	&	2.63	&	3	&	57	&	4.09	&	0	&	100	&	6.25	&	0	&	2.81	\\[-1.0ex]
$S$	&	60	&	100	&	1.24	&	86	&	48	&	1.62	&	20	&	57	&	2.52	&	0	&	100	&	3.68	&	0	&	100	&	6.07	&	0	&	2.39	\\[-1.0ex]
$S_\mathrm{B}$	&	20	&	100	&	1.86	&	113	&	34	&	2.40	&	53	&	44	&	3.71	&	16	&	20	&	5.57	&	1	&	35	&	9.58	&	0	&	6.99	\\[-1.0ex]
$S_\mathrm{B}$	&	20	&	1000	&	1.81	&	117	&	27.2	&	2.34	&	62	&	26.0	&	3.61	&	18	&	14.8	&	5.53	&	1	&	37.3	&	11.47	&	0	&	6.99	\\[-1.0ex]
$S_\mathrm{B}$	&	30	&	100	&	1.65	&	90	&	50	&	2.11	&	54	&	36	&	3.30	&	11	&	31	&	5.11	&	0	&	100	&	7.93	&	0	&	4.38	\\[-1.0ex]
$S_\mathrm{B}$	&	40	&	100	&	1.58	&	39	&	84	&	2.04	&	13	&	88	&	3.18	&	0	&	100	&	5.07	&	0	&	100	&	10.90	&	0	&	2.80	\\[-1.0ex]
$S_\mathrm{B}$	&	50	&	100	&	1.31	&	50	&	81	&	1.68	&	13	&	84	&	2.52	&	0	&	100	&	3.93	&	0	&	100	&	8.75	&	0	&	2.10	\\[-1.0ex]
$S_\mathrm{B}$	&	60	&	100	&	1.07	&	75	&	52	&	1.38	&	24	&	58	&	2.10	&	0	&	100	&	3.17	&	0	&	100	&	4.78	&	0	&	1.91	\\[-1.0ex]
$F_{64}$	&	20	&	100	&	1.80	&	129	&	22	&	2.32	&	66	&	24	&	3.62	&	19	&	15	&	5.68	&	2	&	18	&	9.06	&	0	&	6.77	\\[-1.0ex]
$F_{64}$	&	20	&	1000	&	1.82	&	125	&	19.6	&	2.35	&	65	&	22.2	&	3.63	&	19	&	14.4	&	5.56	&	3	&	12.2	&	14.19	&	0	&	6.77	\\[-1.0ex]
$F_{64}$	&	30	&	100	&	1.65	&	111	&	36	&	2.12	&	68	&	22	&	3.29	&	20	&	12	&	5.11	&	5	&	{\bf 4}	&	7.97	&	0	&	5.67	\\[-1.0ex]
$F_{64}$	&	30	&	1000	&	1.69	&	106	&	41.0	&	2.18	&	60	&	32.7	&	3.41	&	19	&	18.3	&	5.19	&	5	&	8.3	&	12.32	&	0	&	5.67	\\[-1.0ex]
$F_{64}$	&	40	&	100	&	1.56	&	114	&	35	&	2.02	&	57	&	30	&	3.12	&	11	&	24	&	4.74	&	0	&	100	&	8.25	&	0	&	3.82	\\[-1.0ex]
$F_{64}$	&	50	&	100	&	1.40	&	101	&	44	&	1.80	&	50	&	40	&	2.88	&	23	&	15	&	4.43	&	1	&	18	&	6.00	&	0	&	4.70	\\[-1.0ex]
$F_{64}$	&	60	&	100	&	1.20	&	112	&	37	&	1.55	&	77	&	18	&	2.38	&	29	&	8	&	3.50	&	3	&	7	&	5.37	&	0	&	3.70	\\[-1.0ex]
$F_{256}$	&	20	&	100	&	1.81	&	102	&	46	&	2.31	&	48	&	47	&	3.55	&	12	&	34	&	5.35	&	2	&	21	&	8.65	&	0	&	6.62	\\[-1.0ex]
$F_{256}$	&	20	&	1000	&	1.83	&	100	&	48.5	&	2.36	&	46	&	53.8	&	3.67	&	11	&	38.2	&	5.65	&	1	&	34.0	&	12.24	&	0	&	6.62	\\[-1.0ex]
$F_{256}$	&	30	&	100	&	1.67	&	97	&	49	&	2.15	&	55	&	37	&	3.31	&	13	&	25	&	4.93	&	1	&	30	&	10.28	&	0	&	5.35	\\[-1.0ex]
$F_{256}$	&	30	&	1000	&	1.69	&	93	&	51.2	&	2.18	&	52	&	40.9	&	3.36	&	12	&	32.6	&	5.07	&	1	&	27.6	&	10.24	&	0	&	5.35	\\[-1.0ex]
$F_{256}$	&	40	&	100	&	1.49	&	103	&	45	&	1.91	&	50	&	44	&	3.05	&	4	&	49	&	5.05	&	0	&	100	&	8.61	&	0	&	3.44	\\[-1.0ex]
$F_{256}$	&	50	&	100	&	1.37	&	109	&	41	&	1.78	&	48	&	41	&	2.83	&	14	&	25	&	4.14	&	1	&	21	&	6.76	&	0	&	4.31	\\[-1.0ex]
$F_{256}$	&	60	&	100	&	1.13	&	140	&	30	&	1.49	&	63	&	32	&	2.36	&	29	&	16	&	3.61	&	0	&	100	&	6.13	&	0	&	3.50	\\[-1.0ex]
$F_{1024}$	&	20	&	100	&	1.78	&	74	&	80	&	2.28	&	43	&	62	&	3.51	&	7	&	63	&	5.23	&	1	&	39	&	10.10	&	0	&	6.10	\\[-1.0ex]
$F_{1024}$	&	20	&	1000	&	1.81	&	71	&	83.9	&	2.33	&	41	&	64.9	&	3.61	&	7	&	61.7	&	5.58	&	1	&	34.8	&	11.48	&	0	&	6.10	\\[-1.0ex]
$F_{1024}$	&	30	&	100	&	1.69	&	64	&	81	&	2.17	&	42	&	59	&	3.40	&	4	&	61	&	5.36	&	0	&	100	&	9.26	&	0	&	4.31	\\[-1.0ex]
$F_{1024}$	&	40	&	100	&	1.56	&	47	&	78	&	2.02	&	9	&	92	&	3.14	&	0	&	100	&	4.82	&	0	&	100	&	8.13	&	0	&	2.62	\\[-1.0ex]
$F_{1024}$	&	50	&	100	&	1.38	&	82	&	51	&	1.78	&	49	&	35	&	2.80	&	8	&	27	&	4.21	&	0	&	100	&	6.97	&	0	&	3.88	\\[-1.0ex]
$F_{1024}$	&	60	&	100	&	1.15	&	113	&	38	&	1.47	&	76	&	23	&	2.27	&	22	&	16	&	3.44	&	0	&	100	&	5.05	&	0	&	2.86	\\[-1.0ex]
$F_{64\mathrm{,B}}$	&	20	&	100	&	1.80	&	121	&	22	&	2.33	&	71	&	13	&	3.65	&	29	&	{\bf 3}	&	5.67	&	5	&	6	&	10.11	&	0	&	7.51	\\[-1.0ex]
$F_{64\mathrm{,B}}$	&	20	&	1000	&	1.81	&	119	&	26.0	&	2.33	&	71	&	17.0	&	3.60	&	29	&	{\bf 3.4}	&	5.54	&	7	&	{\bf 2.9}	&	12.96	&	0	&	7.51	\\[-1.0ex]
$F_{64\mathrm{,B}}$	&	30	&	100	&	1.71	&	121	&	30	&	2.20	&	81	&	16	&	3.39	&	32	&	{\bf 5}	&	5.32	&	7	&	{\bf 4}	&	8.56	&	0	&	6.69	\\[-1.0ex]
$F_{64\mathrm{,B}}$	&	30	&	1000	&	1.69	&	122	&	29.1	&	2.18	&	81	&	14.9	&	3.39	&	31	&	6.4	&	5.22	&	10	&	{\bf 2.2}	&	10.29	&	0	&	6.69	\\[-1.0ex]
$F_{64\mathrm{,B}}$	&	40	&	100	&	1.49	&	161	&	18	&	1.92	&	107	&	12	&	2.97	&	27	&	10	&	4.57	&	1	&	23	&	6.89	&	0	&	5.30	\\[-1.0ex]
$F_{64\mathrm{,B}}$	&	40	&	1000	&	1.52	&	157	&	14.8	&	1.95	&	104	&	10.0	&	3.03	&	26	&	12.9	&	4.75	&	1	&	19.9	&	10.57	&	0	&	5.30	\\[-1.0ex]
$F_{64\mathrm{,B}}$	&	50	&	100	&	1.42	&	139	&	25	&	1.85	&	72	&	29	&	2.96	&	8	&	36	&	4.78	&	0	&	100	&	7.38	&	0	&	4.65	\\[-1.0ex]
$F_{64\mathrm{,B}}$	&	60	&	100	&	1.16	&	144	&	26	&	1.51	&	68	&	27	&	2.33	&	11	&	28	&	3.50	&	0	&	100	&	5.80	&	0	&	2.83	\\[-1.0ex]
$F_{256\mathrm{,B}}$	&	20	&	100	&	1.80	&	99	&	47	&	2.29	&	62	&	26	&	3.43	&	30	&	{\bf 1}	&	5.09	&	10	&	{\bf 1}	&	7.69	&	2	&	8.98	\\[-1.0ex]
$F_{256\mathrm{,B}}$	&	20	&	1000	&	1.81	&	98	&	50.7	&	2.34	&	58	&	31.6	&	3.65	&	25	&	5.7	&	5.57	&	8	&	{\bf 2.6}	&	11.01	&	0	&	8.98	\\[-1.0ex]
$F_{256\mathrm{,B}}$	&	30	&	100	&	1.71	&	107	&	42	&	2.22	&	66	&	28	&	3.56	&	27	&	10	&	5.64	&	3	&	11	&	9.56	&	0	&	6.55	\\[-1.0ex]
$F_{256\mathrm{,B}}$	&	40	&	100	&	1.56	&	106	&	40	&	2.02	&	73	&	26	&	3.21	&	9	&	28	&	5.39	&	0	&	100	&	8.73	&	0	&	4.34	\\[-1.0ex]
$F_{256\mathrm{,B}}$	&	50	&	100	&	1.39	&	107	&	38	&	1.81	&	42	&	45	&	2.82	&	4	&	45	&	4.66	&	0	&	100	&	9.79	&	0	&	3.45	\\[-1.0ex]
$F_{256\mathrm{,B}}$	&	60	&	100	&	1.11	&	122	&	29	&	1.43	&	49	&	36	&	2.24	&	0	&	100	&	3.79	&	0	&	100	&	7.34	&	0	&	2.07	\\[-1.0ex]
$F_{1024\mathrm{,B}}$	&	20	&	100	&	1.80	&	103	&	45	&	2.32	&	53	&	41	&	3.64	&	15	&	18	&	5.70	&	4	&	8	&	9.36	&	0	&	6.69	\\[-1.0ex]
$F_{1024\mathrm{,B}}$	&	30	&	100	&	1.67	&	83	&	62	&	2.14	&	46	&	50	&	3.24	&	24	&	11	&	4.89	&	2	&	18	&	8.21	&	0	&	5.64	\\[-1.0ex]
$F_{1024\mathrm{,B}}$	&	40	&	100	&	1.58	&	99	&	43	&	2.03	&	41	&	45	&	3.24	&	3	&	48	&	4.92	&	0	&	100	&	8.52	&	0	&	3.60	\\[-1.0ex]
$F_{1024\mathrm{,B}}$	&	50	&	100	&	1.37	&	109	&	38	&	1.77	&	55	&	36	&	2.74	&	4	&	39	&	4.21	&	0	&	100	&	6.50	&	0	&	3.18	\\[-1.0ex]
$F_{1024\mathrm{,B}}$	&	60	&	100	&	1.14	&	135	&	26	&	1.46	&	54	&	33	&	2.22	&	0	&	100	&	3.30	&	0	&	100	&	4.78	&	0	&	2.22	\\
\enddata
\tablenotetext{\ast}{The radii of the patches.\\[-3.5ex]}
\tablenotetext{\star}{The number of random shufflings of the data.\\[-3.5ex]}
\tablenotetext{\dagger}{$AD^\mathrm{s}_{10}$, $AD^\mathrm{s}_{5}$, $AD^\mathrm{s}_{1}$,
and $AD^\mathrm{s}_{0.1}$ delimit the highest 10\,\%, 5\,\%, 1\,\%, and 0.1\,\% of all
$AD^\mathrm{s}$ values from all patches in all randomly shuffled data, respectively.
$AD^\mathrm{s}_0$ is the maximum of all $AD^\mathrm{s}$ values in all patches of all shuffled data.\\[-3.5ex]}
\tablenotetext{\ddagger}{$N^\mathrm{m}_\mathrm{i}$ is the number of patches in the
measured data for which $AD^\mathrm{m}>AD^\mathrm{s}_\mathrm{i}$, where i=10, 5, 1, or 0.1.
$N^\mathrm{m}_0$ is the number of patches in the measured data with $AD^\mathrm{m}>AD^\mathrm{s}_\mathrm{0}$.\\[-3.5ex]}
\tablenotetext{\#}{$P^\mathrm{N}_\mathrm{i}$ is the probability of finding at least
$N^\mathrm{m}_\mathrm{i}$ number of patches with $AD^\mathrm{s}>AD^\mathrm{s}_\mathrm{i}$ in
the randomly shuffled data, where i=10, 5, 1, or 0.1. The cases with
$P^\mathrm{N}_\mathrm{i}\leq 5$\,\% are emphasized in boldface.\\[-3.5ex]}
\tablenotetext{\S}{$AD^\mathrm{m}_{0}$ is the maximum value of the statistic in the measured data.\\[-3.5ex]}
\end{deluxetable*}

\newpage

\begin{deluxetable*}{cccccccccccccccccc}
\tablecaption{\label{tab:results_chi2}
Results using the $\chi^2$ statistic.}
\tablewidth{0pt}
\tabletypesize{\scriptsize}
\tablehead{
\colhead{Tested}	&	\colhead{$r$\tablenotemark{$\ast$}}	&	\colhead{$n$\tablenotemark{$\star$}}	&	\colhead{$\chi_{10}^\mathrm{2\,s}$\tablenotemark{$\dagger$}}	&	\colhead{$N^\mathrm{m}_{10}$\tablenotemark{$\ddagger$}}	&	\colhead{$P^\mathrm{N}_{10}$\tablenotemark{$\#$}}	&	\colhead{$\chi_{5}^\mathrm{2\,s}$\tablenotemark{$\dagger$}}	&	\colhead{$N^\mathrm{m}_{5}$\tablenotemark{$\ddagger$}}	&	\colhead{$P^\mathrm{N}_{5}$\tablenotemark{$\#$}}	&	\colhead{$\chi_{1}^\mathrm{2\,s}$\tablenotemark{$\dagger$}}	&	\colhead{$N^\mathrm{m}_{1}$\tablenotemark{$\ddagger$}}	&	\colhead{$P^\mathrm{N}_{1}$\tablenotemark{$\#$}}	&	\colhead{$\chi_{0.1}^\mathrm{2\,s}$\tablenotemark{$\dagger$}}	&	\colhead{$N^\mathrm{m}_{0.1}$\tablenotemark{$\ddagger$}}	&	\colhead{$P^\mathrm{N}_{0.1}$\tablenotemark{$\#$}}	&
\colhead{$\chi_{0}^\mathrm{2\,s}$\tablenotemark{$\dagger$}}	&	\colhead{$N^\mathrm{m}_0$\tablenotemark{$\ddagger$}}	&	\colhead{$\chi_{0}^\mathrm{2\,m}$\tablenotemark{\S}}
\\[-0.5ex]
\colhead{quantity}	&	\colhead{($^\circ$)}	&	\colhead{}	&	\colhead{}	&	\colhead{}	&	\colhead{(\%)}	&	\colhead{}	&	\colhead{}	&	\colhead{(\%)}	&	\colhead{}	&	\colhead{}	&	\colhead{(\%)}	&	\colhead{}	&	\colhead{}	&	\colhead{(\%)}	&	\colhead{}	&	\colhead{}	&	\colhead{}
}
\startdata
$T_{90}$	&	20	&	100	&	14.36	&	73	&	94	&	18.19	&	34	&	87	&	26.52	&	0	&	100	&	38.71	&	0	&	100	&	72.21	&	0	&	25.23	\\[-1.0ex]
$T_{90}$	&	30	&	100	&	12.41	&	65	&	82	&	14.45	&	26	&	82	&	18.96	&	0	&	100	&	25.98	&	0	&	100	&	46.44	&	0	&	17.91	\\[-1.0ex]
$T_{90}$	&	40	&	100	&	10.84	&	34	&	94	&	12.53	&	9	&	94	&	16.10	&	0	&	100	&	20.73	&	0	&	100	&	31.63	&	0	&	13.88	\\[-1.0ex]
$T_{90}$	&	50	&	100	&	9.67	&	12	&	96	&	11.11	&	1	&	98	&	14.13	&	0	&	100	&	17.79	&	0	&	100	&	26.37	&	0	&	11.22	\\[-1.0ex]
$T_{90}$	&	60	&	100	&	8.20	&	22	&	95	&	9.33	&	10	&	88	&	11.60	&	0	&	100	&	14.70	&	0	&	100	&	19.08	&	0	&	10.50	\\[-1.0ex]
$S$	&	20	&	100	&	12.33	&	86	&	78	&	16.47	&	49	&	51	&	23.59	&	4	&	82	&	31.53	&	0	&	100	&	47.54	&	0	&	26.45	\\[-1.0ex]
$S$	&	20	&	1000	&	12.36	&	86	&	69.9	&	16.40	&	49	&	48.6	&	24.16	&	4	&	78.2	&	32.75	&	0	&	100.0	&	57.12	&	0	&	26.45	\\[-1.0ex]
$S$	&	30	&	100	&	11.47	&	75	&	75	&	13.49	&	40	&	60	&	17.82	&	1	&	90	&	23.69	&	0	&	100	&	37.34	&	0	&	17.93	\\[-1.0ex]
$S$	&	40	&	100	&	10.07	&	38	&	92	&	11.65	&	17	&	86	&	15.01	&	3	&	67	&	20.04	&	0	&	100	&	28.36	&	0	&	18.41	\\[-1.0ex]
$S$	&	50	&	100	&	8.84	&	49	&	81	&	10.15	&	11	&	86	&	12.95	&	0	&	100	&	17.14	&	0	&	100	&	23.64	&	0	&	12.31	\\[-1.0ex]
$S$	&	60	&	100	&	7.81	&	36	&	78	&	8.98	&	16	&	70	&	11.56	&	0	&	100	&	15.45	&	0	&	100	&	22.07	&	0	&	11.24	\\[-1.0ex]
$S_\mathrm{B}$	&	20	&	100	&	12.14	&	99	&	51	&	15.84	&	56	&	32	&	24.21	&	7	&	57	&	32.18	&	0	&	100	&	46.75	&	0	&	27.66	\\[-1.0ex]
$S_\mathrm{B}$	&	20	&	1000	&	12.29	&	98	&	52.4	&	16.09	&	54	&	34.3	&	24.21	&	7	&	59.6	&	32.46	&	0	&	100.0	&	54.78	&	0	&	27.66	\\[-1.0ex]
$S_\mathrm{B}$	&	30	&	100	&	11.31	&	78	&	77	&	13.25	&	49	&	49	&	17.62	&	12	&	40	&	22.97	&	0	&	100	&	31.47	&	0	&	22.91	\\[-1.0ex]
$S_\mathrm{B}$	&	40	&	100	&	10.21	&	64	&	71	&	11.87	&	29	&	68	&	15.38	&	5	&	52	&	22.19	&	0	&	100	&	30.88	&	0	&	16.85	\\[-1.0ex]
$S_\mathrm{B}$	&	50	&	100	&	8.76	&	47	&	82	&	10.09	&	19	&	81	&	12.66	&	0	&	100	&	15.76	&	0	&	100	&	21.94	&	0	&	12.51	\\[-1.0ex]
$S_\mathrm{B}$	&	60	&	100	&	7.56	&	37	&	83	&	8.72	&	4	&	90	&	11.15	&	0	&	100	&	14.29	&	0	&	100	&	18.90	&	0	&	9.15	\\[-1.0ex]
$F_{64}$	&	20	&	100	&	13.74	&	100	&	50	&	18.36	&	49	&	51	&	25.56	&	4	&	72	&	35.47	&	0	&	100	&	47.58	&	0	&	31.03	\\[-1.0ex]
$F_{64}$	&	20	&	1000	&	14.12	&	97	&	55.9	&	18.60	&	43	&	69.5	&	25.93	&	3	&	78.3	&	37.04	&	0	&	100.0	&	67.73	&	0	&	31.03	\\[-1.0ex]
$F_{64}$	&	30	&	100	&	11.17	&	48	&	98	&	13.36	&	1	&	100	&	17.92	&	0	&	100	&	23.27	&	0	&	100	&	37.76	&	0	&	15.18	\\[-1.0ex]
$F_{64}$	&	30	&	1000	&	11.25	&	47	&	96.7	&	13.46	&	1	&	100.0	&	18.19	&	0	&	100.0	&	24.32	&	0	&	100.0	&	43.52	&	0	&	15.18	\\[-1.0ex]
$F_{64}$	&	40	&	100	&	9.63	&	16	&	98	&	11.30	&	0	&	100	&	14.79	&	0	&	100	&	19.27	&	0	&	100	&	25.77	&	0	&	10.79	\\[-1.0ex]
$F_{64}$	&	50	&	100	&	8.16	&	33	&	91	&	9.47	&	3	&	98	&	12.05	&	0	&	100	&	15.49	&	0	&	100	&	20.89	&	0	&	10.06	\\[-1.0ex]
$F_{64}$	&	60	&	100	&	7.22	&	46	&	72	&	8.29	&	18	&	69	&	10.61	&	0	&	100	&	13.93	&	0	&	100	&	18.45	&	0	&	9.42	\\[-1.0ex]
$F_{256}$	&	20	&	100	&	15.18	&	102	&	48	&	19.75	&	36	&	84	&	27.21	&	3	&	86	&	39.02	&	0	&	100	&	52.37	&	0	&	32.76	\\[-1.0ex]
$F_{256}$	&	20	&	1000	&	15.18	&	102	&	44.0	&	19.59	&	38	&	80.8	&	27.34	&	3	&	80.1	&	39.09	&	0	&	100.0	&	70.76	&	0	&	32.76	\\[-1.0ex]
$F_{256}$	&	30	&	100	&	12.29	&	89	&	63	&	14.50	&	21	&	86	&	18.75	&	0	&	100	&	24.56	&	0	&	100	&	40.75	&	0	&	18.50	\\[-1.0ex]
$F_{256}$	&	30	&	1000	&	12.31	&	87	&	64.6	&	14.55	&	19	&	91.2	&	19.33	&	0	&	100.0	&	26.07	&	0	&	100.0	&	43.41	&	0	&	18.50	\\[-1.0ex]
$F_{256}$	&	40	&	100	&	10.48	&	115	&	37	&	12.20	&	55	&	38	&	15.89	&	3	&	58	&	21.13	&	0	&	100	&	32.99	&	0	&	17.45	\\[-1.0ex]
$F_{256}$	&	50	&	100	&	9.17	&	149	&	22	&	10.57	&	75	&	22	&	13.43	&	2	&	65	&	17.08	&	0	&	100	&	27.19	&	0	&	14.26	\\[-1.0ex]
$F_{256}$	&	60	&	100	&	7.79	&	130	&	32	&	8.87	&	72	&	33	&	11.08	&	9	&	36	&	13.87	&	0	&	100	&	18.15	&	0	&	13.07	\\[-1.0ex]
$F_{1024}$	&	20	&	100	&	12.32	&	69	&	93	&	16.53	&	47	&	55	&	24.45	&	5	&	72	&	33.62	&	0	&	100	&	49.01	&	0	&	26.11	\\[-1.0ex]
$F_{1024}$	&	20	&	1000	&	12.47	&	69	&	93.8	&	16.65	&	47	&	57.5	&	24.97	&	4	&	76.2	&	34.41	&	0	&	100.0	&	72.58	&	0	&	26.11	\\[-1.0ex]
$F_{1024}$	&	30	&	100	&	11.40	&	36	&	100	&	13.54	&	8	&	99	&	18.11	&	0	&	100	&	25.29	&	0	&	100	&	35.62	&	0	&	17.67	\\[-1.0ex]
$F_{1024}$	&	40	&	100	&	9.97	&	8	&	100	&	11.57	&	0	&	100	&	14.97	&	0	&	100	&	18.96	&	0	&	100	&	24.45	&	0	&	11.13	\\[-1.0ex]
$F_{1024}$	&	50	&	100	&	8.68	&	12	&	98	&	10.01	&	2	&	96	&	12.87	&	0	&	100	&	17.03	&	0	&	100	&	23.33	&	0	&	10.20	\\[-1.0ex]
$F_{1024}$	&	60	&	100	&	7.45	&	18	&	93	&	8.63	&	4	&	93	&	11.00	&	0	&	100	&	14.31	&	0	&	100	&	21.06	&	0	&	10.57	\\[-1.0ex]
$F_{64\mathrm{,B}}$	&	20	&	100	&	15.89	&	107	&	35	&	20.10	&	59	&	24	&	28.89	&	16	&	23	&	44.08	&	1	&	23	&	64.67	&	0	&	44.28	\\[-1.0ex]
$F_{64\mathrm{,B}}$	&	20	&	1000	&	16.05	&	106	&	34.1	&	20.24	&	58	&	29.3	&	28.73	&	16	&	22.0	&	41.33	&	1	&	22.9	&	73.94	&	0	&	44.28	\\[-1.0ex]
$F_{64\mathrm{,B}}$	&	30	&	100	&	11.89	&	138	&	7	&	14.16	&	61	&	32	&	19.17	&	9	&	38	&	24.83	&	4	&	9	&	34.47	&	0	&	27.64	\\[-1.0ex]
$F_{64\mathrm{,B}}$	&	30	&	1000	&	12.10	&	128	&	20.2	&	14.48	&	54	&	40.6	&	19.52	&	8	&	44.7	&	26.50	&	3	&	10.0	&	48.26	&	0	&	27.64	\\[-1.0ex]
$F_{64\mathrm{,B}}$	&	40	&	100	&	9.96	&	157	&	13	&	11.46	&	66	&	29	&	14.79	&	4	&	59	&	19.54	&	0	&	100	&	32.74	&	0	&	15.22	\\[-1.0ex]
$F_{64\mathrm{,B}}$	&	40	&	1000	&	10.13	&	147	&	15.4	&	11.77	&	54	&	39.7	&	15.31	&	0	&	100.0	&	19.79	&	0	&	100.0	&	35.48	&	0	&	15.22	\\[-1.0ex]
$F_{64\mathrm{,B}}$	&	50	&	100	&	8.92	&	100	&	45	&	10.32	&	41	&	47	&	13.17	&	4	&	51	&	16.39	&	0	&	100	&	20.34	&	0	&	14.83	\\[-1.0ex]
$F_{64\mathrm{,B}}$	&	60	&	100	&	7.64	&	74	&	53	&	8.71	&	30	&	53	&	11.16	&	3	&	42	&	14.05	&	0	&	100	&	18.74	&	0	&	12.98	\\[-1.0ex]
$F_{256\mathrm{,B}}$	&	20	&	100	&	16.12	&	112	&	25	&	20.35	&	44	&	64	&	28.42	&	2	&	84	&	41.06	&	0	&	100	&	61.49	&	0	&	35.09	\\[-1.0ex]
$F_{256\mathrm{,B}}$	&	20	&	1000	&	16.13	&	111	&	24.9	&	20.25	&	47	&	57.6	&	28.70	&	2	&	82.2	&	41.84	&	0	&	100.0	&	78.09	&	0	&	35.09	\\[-1.0ex]
$F_{256\mathrm{,B}}$	&	30	&	100	&	12.21	&	80	&	73	&	14.48	&	30	&	76	&	19.21	&	5	&	56	&	24.91	&	0	&	100	&	37.10	&	0	&	20.61	\\[-1.0ex]
$F_{256\mathrm{,B}}$	&	40	&	100	&	10.18	&	44	&	90	&	11.82	&	7	&	98	&	15.33	&	0	&	100	&	19.31	&	0	&	100	&	28.05	&	0	&	14.00	\\[-1.0ex]
$F_{256\mathrm{,B}}$	&	50	&	100	&	8.83	&	36	&	90	&	10.13	&	9	&	92	&	12.88	&	0	&	100	&	16.07	&	0	&	100	&	25.01	&	0	&	10.93	\\[-1.0ex]
$F_{256\mathrm{,B}}$	&	60	&	100	&	7.60	&	47	&	73	&	8.62	&	11	&	81	&	10.71	&	1	&	70	&	14.00	&	0	&	100	&	20.33	&	0	&	10.80	\\[-1.0ex]
$F_{1024\mathrm{,B}}$	&	20	&	100	&	16.72	&	97	&	56	&	20.51	&	51	&	45	&	28.33	&	9	&	51	&	39.16	&	0	&	100	&	66.10	&	0	&	37.13	\\[-1.0ex]
$F_{1024\mathrm{,B}}$	&	30	&	100	&	12.02	&	102	&	40	&	14.27	&	45	&	59	&	18.92	&	3	&	67	&	25.91	&	0	&	100	&	44.74	&	0	&	20.71	\\[-1.0ex]
$F_{1024\mathrm{,B}}$	&	40	&	100	&	10.30	&	69	&	76	&	12.07	&	25	&	77	&	16.08	&	5	&	54	&	21.27	&	2	&	13	&	27.94	&	0	&	25.26	\\[-1.0ex]
$F_{1024\mathrm{,B}}$	&	50	&	100	&	8.75	&	99	&	48	&	10.04	&	64	&	33	&	12.76	&	15	&	21	&	16.05	&	2	&	19	&	20.57	&	0	&	17.48	\\[-1.0ex]
$F_{1024\mathrm{,B}}$	&	60	&	100	&	7.35	&	117	&	31	&	8.42	&	72	&	26	&	10.67	&	23	&	13	&	13.59	&	2	&	13	&	16.68	&	0	&	14.71	\\
\enddata
\tablenotetext{\ast}{The radii of the patches.\\[-3.5ex]}
\tablenotetext{\star}{The number of random shufflings of the data.\\[-3.5ex]}
\tablenotetext{\dagger}{$\chi^\mathrm{2\,s}_{10}$, $\chi^\mathrm{2\,s}_{5}$, $\chi^\mathrm{2\,s}_{1}$,
and $\chi^\mathrm{2\,s}_{0.1}$ delimit the highest 10\,\%, 5\,\%, 1\,\%, and 0.1\,\% of all
$\chi^\mathrm{2\,s}$ values from all patches in all randomly shuffled data, respectively.
$\chi^\mathrm{2\,s}_{0}$ is the maximum of all $\chi^\mathrm{2\,s}$ values in all patches of all shuffled data.\\[-3.5ex]}
\tablenotetext{\ddagger}{$N^\mathrm{m}_\mathrm{i}$ is number of patches in the
measured data for which $\chi^\mathrm{2\,m}>\chi^\mathrm{2\,s}_\mathrm{i}$, where i=10, 5, 1, 0.1.
$N^\mathrm{m}_0$ is number of patches in the measured data with $\chi^\mathrm{2\,m}>\chi^\mathrm{2\,s}_{0}$.\\[-3.5ex]}
\tablenotetext{\#}{$P^\mathrm{N}_\mathrm{i}$ is the probability of finding at least
$N^\mathrm{m}_\mathrm{i}$ number of patches with $\chi^\mathrm{2\,s}>\chi^\mathrm{2\,s}_\mathrm{i}$
in the randomly shuffled data, where i=10, 5, 1, or 0.1. The cases with
$P^\mathrm{N}_\mathrm{i}\leq 5$\,\% are emphasized in boldface.\\[-3.5ex]}
\tablenotetext{\S}{$\chi^\mathrm{2\,m}_{0}$ is the maximum value of the statistic in the measured data.\\[-3.5ex]}
\end{deluxetable*}

\end{document}